\PassOptionsToPackage{table,xcdraw}{xcolor}

\documentclass[10pt,conference]{IEEEtran}

\usepackage[numbers]{natbib}
\usepackage{hyperref}

\usepackage{fancybox}
\usepackage{multirow}
\usepackage{flushend}
\usepackage{booktabs}
\usepackage{tabularx}
\usepackage{comment}
\usepackage{array}
\usepackage[flushleft]{threeparttable}
\usepackage{mdframed}

\usepackage{subfig}

\usepackage{listings}
\usepackage{courier}
\usepackage[normalem]{ulem}
\usepackage{color}
\usepackage{hyphenat}
\usepackage{alltt                                    
	, multirow
	,subfig
	, booktabs
	, listings
	,float
	,verbatim
	,mathtools
	,url
	,amsmath
}
\usepackage{framed,lipsum}
\usepackage[most]{tcolorbox}
\usepackage{graphics} 
\usepackage{graphicx}
\usepackage{pgfplots, pgfplotstable}
\pgfplotsset{compat=1.15}
\usepackage{tikz}
\usetikzlibrary{patterns}

\usepackage{algorithm2e}
\usepackage{algpseudocode}

\newcounter{o}
\usepackage{xspace}

\usepackage[export]{adjustbox}

\usepackage{tikz}
\usepackage{pgf-pie}
\usetikzlibrary{positioning,shadows}

\newif\ifpienumberinlegend
\pgfkeys{/number in legend/.code=
    \expandafter\let\expandafter\ifpienumberinlegend
    \csname if#1\endcsname
    \ifpienumberinlegend

    \def\beforenumber##1\afternumber{}%
    \fi,
    /number in legend/.default=true
}

\definecolor{1c1}{RGB}{188,162,6}
\definecolor{1c2}{RGB}{137,129,80}
\definecolor{1c3}{RGB}{239,167,31}
\definecolor{1c4}{RGB}{88,194,241}
\definecolor{1c5}{RGB}{6,180,188}

\tikzset{mynode/.style={draw=white,solid,circle,fill=green,inner sep=1pt, thick,
text=black}}
\tikzset{arrow line/.style={dashed, line width= 2.5pt, color=#1}}

\def\it{\textit}

\usepackage{paralist}

\usepackage{balance}

\normalsize

\usepackage{enumitem}

\usepackage{tikz}

\lstdefinestyle{inlinecode}{basicstyle={\ttfamily\scriptsize\bfseries}}

\newcommand{\urls}[1]{{\scriptsize\url{#1}}}
\usepackage{tcolorbox}

\usepackage{soul}
\usepackage{paralist}
\usepackage[outercaption]{sidecap}    

\usepackage{enumitem}
\usepackage[T1]{fontenc}
\usepackage{pifont}

\definecolor{lightgray}{gray}{.92}

\newtcolorbox
{mybox}[2][]{colbacktitle=red!10!white,
colback=blue!10!white,coltitle=black!70!black,
title={#2},fonttitle=\bfseries,#1}
\usepackage{graphicx}
\usepackage{subfig}

\definecolor{Gray}{gray}{0.9}

\begin{document}


\title{Do Subjectivity and Objectivity Always Agree? A Case Study with Stack Overflow Questions}

\author{\IEEEauthorblockN{Saikat Mondal}
\IEEEauthorblockA{University of Saskatchewan, Canada\\ }
{saikat.mondal@usask.ca}
\and
\IEEEauthorblockN {Mohammad Masudur Rahman}
\IEEEauthorblockA{Dalhousie University, Canada\\ }
{masud.rahman@dal.ca}
\and
\IEEEauthorblockN {Chanchal K. Roy}
\IEEEauthorblockA{University of Saskatchewan, Canada\\ }
{chanchal.roy@usask.ca}
}








\maketitle

\begin{abstract}
In Stack Overflow (SO), the quality of posts (i.e., questions and answers) is subjectively evaluated by users through a voting mechanism. The net votes (upvotes $-$  downvotes) obtained by a post are often considered an approximation of its quality. However, about half of the questions that received working solutions got more downvotes than upvotes. Furthermore, about 18\% of the accepted answers (i.e., verified solutions) also do not score the maximum votes. All these counter-intuitive findings cast doubts on the reliability of the evaluation mechanism employed at SO.
Moreover, many users raise concerns against the evaluation, especially downvotes to their posts. Therefore, rigorous verification of the subjective evaluation is highly warranted to ensure a non-biased and reliable quality assessment mechanism. In this paper, we compare the subjective assessment of questions with their objective assessment using 2.5 million questions and ten text analysis metrics. 
According to our investigation, four objective metrics agree with the subjective evaluation, two do not agree, one either agrees or disagrees, and the remaining three neither agree nor disagree with the subjective evaluation.
We then develop machine learning models to classify the \emph{promoted} and \emph{discouraged} questions. Our models outperform the state-of-the-art models with a maximum of about 76\%--87\% accuracy.
\end{abstract}







\section{Introduction} \label{intro}
SO has emerged as one of the largest and most popular Q\&A sites for programming problems and solutions. It has been a valuable knowledge base containing about 23 million questions and 35 million answers as of December 2022, which can be leveraged for various problem-solving during software development \cite{HDP-Treude-2011}. The quality of any question or answer in SO is subjectively evaluated by the users through a voting mechanism. High-quality posts are generally voted up, whereas vague, unclear, or ambiguous posts are likely to be voted down.
The net votes (upvotes $-$ downvotes) cast against a given post in SO forms an evaluation metric called \emph{score}, which also approximates the subjective quality of the post. However, according to our investigation, 48\% of the questions that received working solutions have a score below zero \cite{api}. That is, they receive more downvotes than upvotes. In other words, although these questions were valid and were answered successfully, they might not be considered of high quality by others. About 18\% of the accepted answers (i.e., verified solutions) also fail to receive the maximum votes among all answers in a Q\&A thread \cite{api}. That is, the best answers might not have been chosen as the solutions. The users of SO are also often disappointed by the evaluation against their questions (e.g., Fig. 1) and answers (e.g., Fig. 2). All these findings above suggest the highly subjective nature of the votes cast by the users in SO. 

Furthermore, several previous studies also show that the incentive systems of Q\&A sites may not always drive certain users in a positive way. For example, opportunistic users may find loopholes, and aggressively game the system for their advantage \cite{WPE-Hsieh-2010, TMI-Jan-2017, HDU-Wang-2018}.
All these findings above cast serious doubts on the reliability of the evaluation mechanism employed at SO. In order to ensure a fair, reliable and effective evaluation mechanism, an investigation is warranted that verifies the subjective evaluation of a post (by the SO users) with the corresponding objective evaluation. Objective evaluation involves numerical metrics that estimate  the quality by minimizing subjective bias.
Such an agreement analysis between subjective evaluation and objective evaluation might help us better understand the potential or peril of the voting systems employed in SO.

\begin{figure}
	\centering
	\includegraphics[width=3.2in]{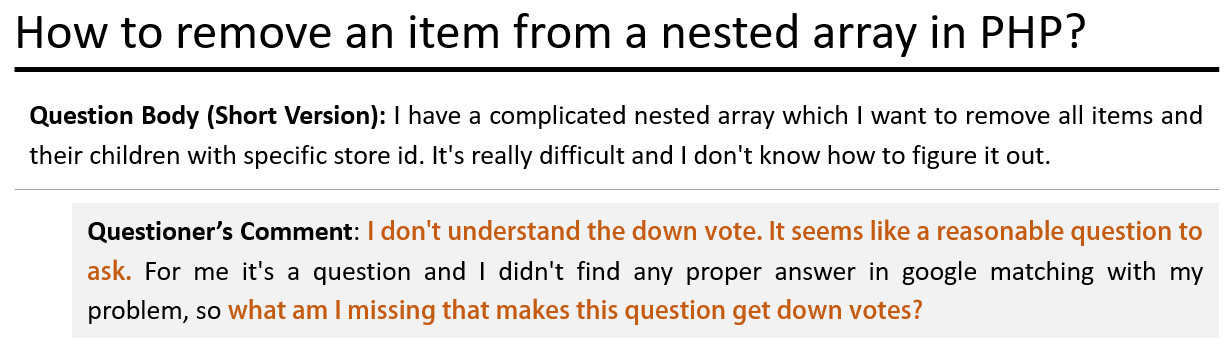}
	\caption{\small{An example of a question submitter complaining in a comment that the subjective evaluation of his question is not reasonable (\url{https://stackoverflow.com/questions/47183288}).}}
	\label{fig:motivating-example-01}
\end{figure}

\begin{figure}
	\centering
	\includegraphics[width=3.2in]{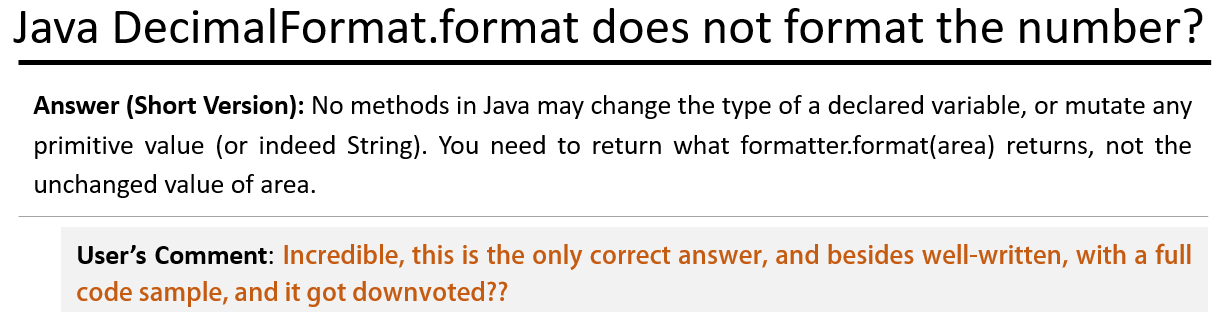}
	\caption{\small{An example of a user complaining in a comment that the answer must not be voted down (\url{https://stackoverflow.com/questions/11765616}).}}
	\label{fig:motivating-example-02}
\end{figure}

Several studies \cite{WMA-Nasehi-2012, USO-Treude-2017, AAO-Ercan-2015, MSA-Calefato-2015} analyze questions and answers of SO and attempt to assess their quality. \citet{WMA-Nasehi-2012} analyze the quality of code snippets included in SO posts, and study the aspects that make up a good code example. \citet{USO-Treude-2017} investigate to what extent developers perceive such code examples as self-explanatory. The appropriateness of code explanation is further studied by \citet{AAO-Ercan-2015} using Recall-Oriented Understudy for Gisting Evaluation (ROUGE) \cite{RUA-Ganesan-2018}. \citet{MSA-Calefato-2015} focus on the four lightweight, user-oriented features (e.g., user reputation) to predict the high-quality answers from technical Q\&A sites. \citet{TDT-Novielli-2014} argue that the emotional style of a technical question could be a quality aspect that might influence the answering time of the question. Thus, the above studies simply rely on the lightweight, ad-hoc attributes to estimate a post's quality. A few other studies \cite{ILQ-Ponzanelli-2014, UAC-Ponzanelli-2014,QQN-Duijn-2015}  attempt to separate up-voted questions from down-voted questions using both the subjective (e.g., user reputation) and objective (e.g., text readability) metrics. While our work overlaps with them in terms of methodology, our research goals are different. In particular, we attempt to verify whether the subjective evaluation makes sense and the high-scored (hereby, \emph{promoted}) questions on SO are actually preferable to the low-scored (hereby, \emph{discouraged}) ones in terms of different objective quality metrics.

In this study, we conduct a case study with 2.5 million questions of SO where we verify the subjective quality of SO questions with objective evaluation. In particular, we check whether the quality of questions perceived by users matches the objective quality estimate. 
We have several findings from this agreement analysis. We found that \emph{four} of the metrics totally agree,  \emph{one} metric somewhat agrees, and \emph{two} metrics totally disagree with the subjective evaluation of the quality of a post. We also develop classification models based on objective measures that outperforms the state-of-the-art classification models. 
SO might incorporate a mechanism to estimate the objective quality of posts using our metrics. Then users could assess the quality of the posts at SO and also their post quality during submission time. Alternatively, a client-side browser plug-in can be developed employing our model. By installing the plug-in, users can estimate the objective quality of not only their existing posts but also their new posts.
In this paper, we answer two research questions and thus make two contributions as follows: 

\smallskip
\textbf{(RQ$\mathbf{_1})$} \textbf{Does crowd-based subjective assessment of the quality of SO questions agree with the objective assessment of their quality?} We conduct a comparative analysis between the promoted and discouraged questions using \emph{ten} objective metrics from the literature. Our analysis reports several findings.  
First, topic entropy, metric entropy, text-code ratio, and text-code correlation support the subjective evaluation. 
That is, promoted questions are specific, less ambiguous, use unusual terms, and their code snippets are well explained with accompanying prose. 
Second, the reusability of the code snippet and text readability do not agree with subjective evaluation. 
That is, the code examples included in promoted questions are hard to reuse. 
They are less parsable than those of the discouraged examples. Code examples included in discouraged questions are complete and more parsable. However, they often have redundant statements that might cost additional time in code analysis. According to our investigation, promoted questions often contain program elements (e.g., API classes) within the texts. The existing readability metrics consider such elements as complex words that hurt readability.
Third, code understandability could be either higher or lower for the promoted questions based on the programming languages. 
Fourth, title quality, code readability, and sentiment polarity are almost similar for both question types.

\smallskip
\textbf{(RQ$\mathbf{_2})$} \textbf{Can we classify the promoted and discouraged questions of SO Q\&A site?} 
We develop five machine learning models to evaluate the quality of questions and to identify promoted and discouraged questions at creation time. 
We analyze the features and rank them. Our analysis reports that topic entropy, readability of text, metric entropy, and text-code ratio are the top four strong predictors.
Our models classify the questions with a maximum of about 76\%$-$87\% accuracy that outperforms the accuracy of the state-of-the-art \cite{UAC-Ponzanelli-2014} from the literature.
Predicting a potentially down-voted question in advance might greatly help one improve the question during its submission.



\begin{figure}
	\centering
	\includegraphics[width=3.45in]{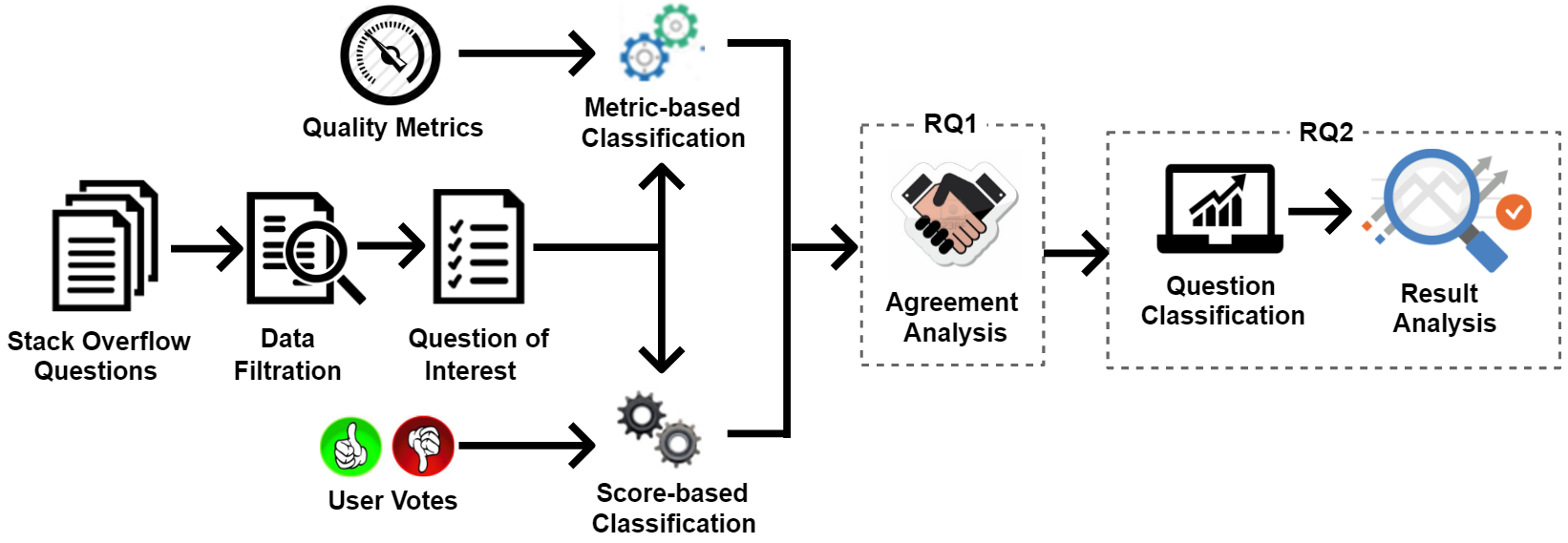}
	\caption{Schematic diagram of our study.}
	\label{fig:system-diagram}
\end{figure}

\smallskip
\noindent\textbf{Replication Package} of this study can be found in our online appendix \cite{replicationPackage}.

\section{Study Methodology}\label{methodology}
\label{methodilogy}

Fig. \ref{fig:system-diagram} shows the schematic diagram of our study. We first collect about 2.5 million questions from Stack Overflow. We collect these questions from four popular programming languages employing three restrictions. Then we use ten objective metrics to assess their quality from different aspects. Next, we analyze the agreement between subjective and objective quality measures. Finally, we classify the promoted and discouraged questions based on their objective quality using five machine learning models and evaluate the models' performance.
The following sections discuss different steps of our methodology.

\begin{table}[!ht]
	\centering
	\captionsetup{justification=centering, labelsep=newline}
	\caption{Summary of the Study Dataset}
	\label{table:total-data-summary}
	
	\resizebox{3.3in}{!}{%
	\begin{tabular}{l|c|c|c} \hline
		\textbf{Topic} & \textbf{Promoted Questions}   & \textbf{Discouraged Questions} & \textbf{Total}\\ \hline \hline
		C\# &  550,173 & 76,181 & 626,354 \\ \hline
		Java & 585,410  & 116,268 & 701,678 \\ \hline
		Javascript & 598,670  & 114,645 & 713,315 \\ \hline
		Python & 418,700  & 67,035 & 485,735 \\ \hline  \hline
		\textbf{Total} & 2,152,953  & 374,129 & \textbf{2,527,082} \\ \hline
	\end{tabular}
	}
	
\end{table}

\section{Data Collection}
\label{dataCollection}

To verify the subjective assessment of SO questions with the objective evaluation, we collect a total number of 2,527,082 questions from SO using StackExchange Data API \cite{api}. In particular, we collect questions related to four major programming languages: C\#, Java, JavaScript, and Python. We choose these questions under certain restrictions -- \textbf{(i)} each question must have at least one answer to ensure that the question has attracted the attention of the community \textbf{(ii)} the question has a score not equal to zero, which indicates that it has obtained at least one up or down vote, and \textbf{(iii)} questions are posted in SO in the year 2017 or earlier which suggests that the questions have been assessed by SO users for a significant time period. Table \ref{table:total-data-summary} lists the dataset for our study. We categorize the questions into \emph{two} sub-categories according to their score as follows.

\begin{itemize}
	
	\item\textbf{Promoted questions}: Questions with a score greater than 0 (i.e., questions with more upvotes).
	\item\textbf{Discouraged questions}: Questions with a score less than 0 (i.e., questions with more downvotes).
	
\end{itemize}

Several objective measures (e.g., reusability of code snippets) can be applied to the code snippet included with questions. According to our investigation, 1,909,539 out of 2,527,082 (i.e., 75.56\%) questions have at least one line of code. We identify such questions and extract code snippets using specialized HTML tags such as \texttt{<code>} under \texttt{<pre>}, and select them for our study. We also collect a total of $54,463$ tags from the \emph{Tags} table using StackExchange Data API \cite{api}. We use this data in calculating the topic entropy of a question.

\section{A comparative study between promoted and discouraged questions using ten objective metrics (RQ$_1$)}
\label{comparativeStudy}

A SO question has three components: title, body, and tags. The body often contains code snippets along with textual descriptions. To answer {RQ$_1$}, we analyze both promoted and discouraged questions using \emph{ten} metrics from the literature to quantify the objective quality of these three components.
The metrics are -- (a) title quality, (b) text readability, (c) code readability, (d) text-code ratio, (e) text-code correlation, (f) code reusability, (g) code understandability, (h) topic entropy, (i) metric entropy, and (j) sentiment polarity. All the reported metrics can be calculated by considering the data available during the question submission. 
Title quality estimates the integrity between the title and other parts of a question. Explanation quality metrics (e.g., text-code ratio, text-code correlation) estimate the comprehensiveness of the code explanation and the coherence between code segments and their explanation. Readability metrics (e.g., text and code readability) estimate the readability of texts and code snippets. Topic and metric entropy estimate the ambiguity of the question topics and the randomness of the terms used in the question texts. Code usability metrics (e.g., code reusability and understandability) estimate the parsability and understandability of code snippets. Finally, the sentiment polarity metric estimates the positive, negative, or neutral sentiment from the question text.
In particular, we compare the objective quality metrics of promoted questions to that of discouraged questions. Our goals are to determine (1) whether there exist any noticeable differences between these two sets of metrics, as a result, (2) whether they either support or contradict the subjective assessment made by SO users. Finally, we attempt to derive a conclusion from such findings for {RQ$_1$}. Our comparative study is divided into several different analyses as follows:

\subsection{Title Quality (TQ)}
\label{sec-tq}
The title is an important part that should properly summarize the question, whereas the body texts explain the programming problem in detail. However, poorly written titles might fail to draw the attention of the potential users who could have answered the question. We thus measure the title quality using a popular, lightweight metric, namely ROUGE. ROUGE automatically determines the quality of a summary by comparing it with human-written summaries. ROUGE has a high correlation with human scores \cite{AAO-Ercan-2015}. We use ROUGE 2.0 \cite{RUA-Ganesan-2018} for our analysis. We explored different types of ROUGE metrics such as ROUGE-N (N-gram Co-Occurrence Statistics), ROUGE-L (Longest Common Subsequence), ROUGE-W (Weighted Longest Common Subsequence), ROUGE-S (Skip-Bigram Co-Occurrence Statistics), and ROUGE-SU (Skip-Bigram with the addition of unigram as counting unit). After a careful analysis, we find that ROUGE-1 (unigram) works reasonably well in our problem context. We thus use ROUGE-1 to assess title quality. Here, we consider the title of a question as a reference summary (i.e., human-generated summary). ROUGE takes the body texts that contain more detailed information about the problem as a system summary. Then we calculate recall to determine the quality of the title as follows.

\begin{figure}[htb]
	
	\centering
	
	\subfloat[For each of the programming languages]
	{\includegraphics[width=1.6in]{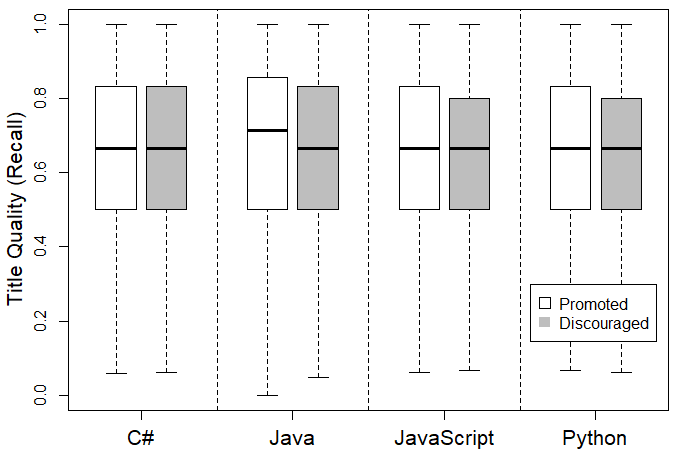}\label{fig:title-quality-programming-languages(a)}}
	\hspace{2mm}\subfloat[For all the programming languages]
	{\includegraphics[width=1.67in]{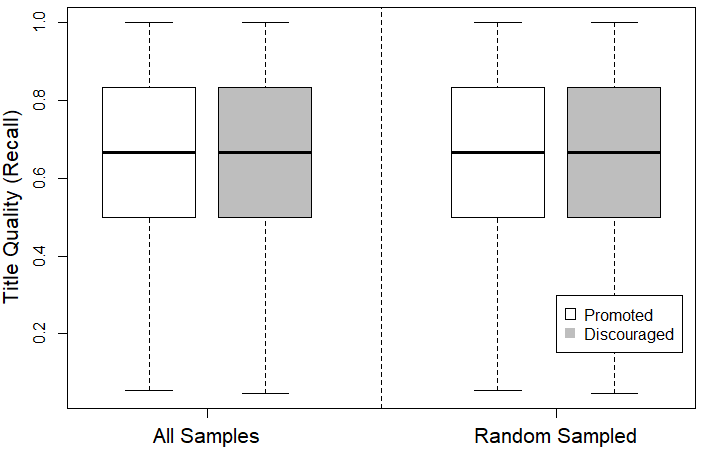}\label{fig:title-quality-all(b)}}
	
	\caption{Title quality}
	\label{fig:title-quality-bar-plot}
	
\end{figure}

\smallskip
    $Recall=\frac{{COUNT}(\text{Reference Summary} \cap \text{System Summary)}}{{COUNT}(\text{Words in Reference Summary)}}$
\vspace{2mm}

\textbf{RQ\textsubscript{1}(a): Does the title quality agree with subjective evaluation?} As shown in Fig. \ref{fig:title-quality-bar-plot}, we did not find any significant differences in the title quality between promoted and discouraged questions. First, we attempt to determine the title quality of both the promoted and discouraged questions. We find that the titles of Java-related promoted questions have slightly higher recall than those of discouraged questions. The other three languages have a nearly identical recall for both question categories. Then we combined the samples of all the languages and also did not notice any significant differences. Finally, we undersampled the promoted questions to balance the dataset and observed similar results.

	
	

\begin{figure*}
	\centering
 
	\subfloat[\footnotesize{C\# questions with no code snippet}]
	{\includegraphics[width=1.68in]{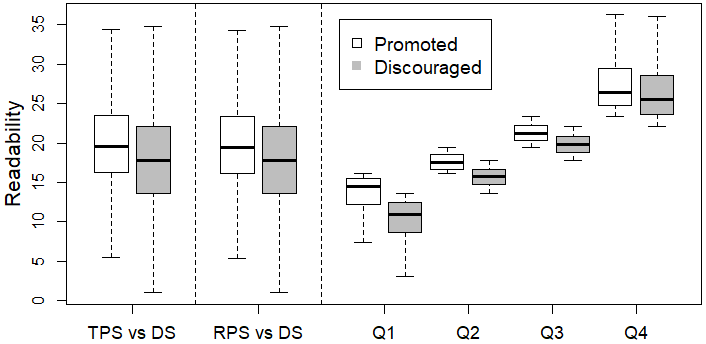}\label{fig:JavaTextReadabilityBox(a1)}}
	\hspace{2mm}\subfloat[C\# questions with code snippet]
	{\includegraphics[width=1.68in]{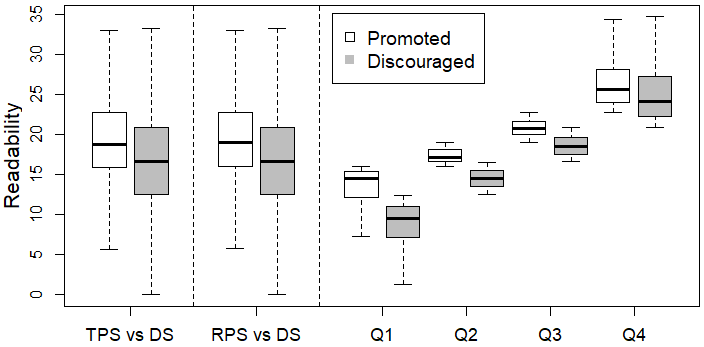}\label{fig:JavaTextReadabilityBox(a2)}}
	\hspace{2mm}\subfloat[Java questions with no code snippet]
	{\includegraphics[width=1.68in]{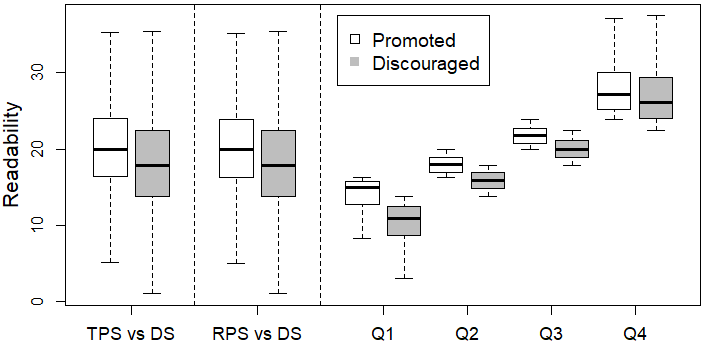}\label{fig:CSharpTextReadabilityBox(b1)}}
	\hspace{2mm}\subfloat[Java questions with code snippet]
	{\includegraphics[width=1.68in]{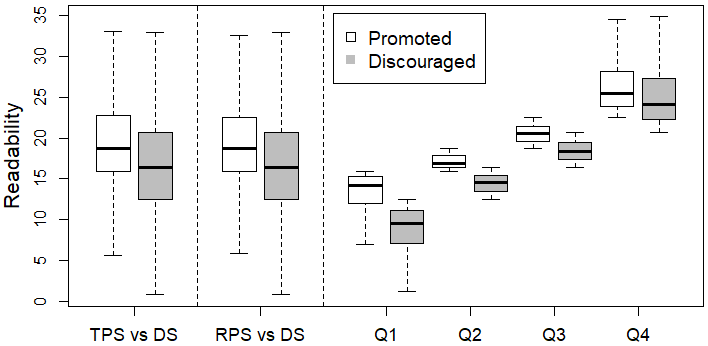}\label{fig:CSharpTextReadabilityBox(b2)}}

    \vspace{-2mm}
	\subfloat[Javascript questions with no code snippet]
	{\includegraphics[width=1.68in]{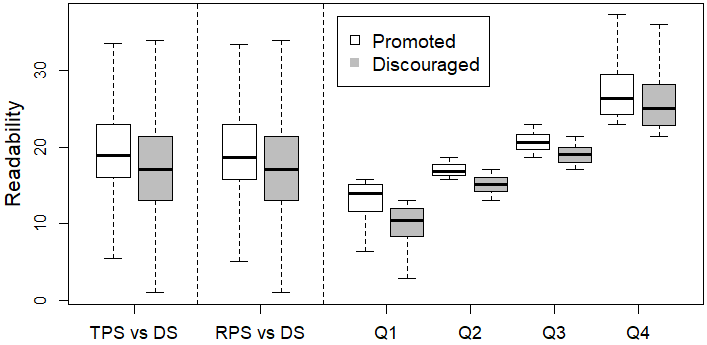}\label{fig:JavascriptTextReadabilityBox(c1)}}
	\hspace{2mm}\subfloat[Javascript questions with code snippet]
	{\includegraphics[width=1.68in]{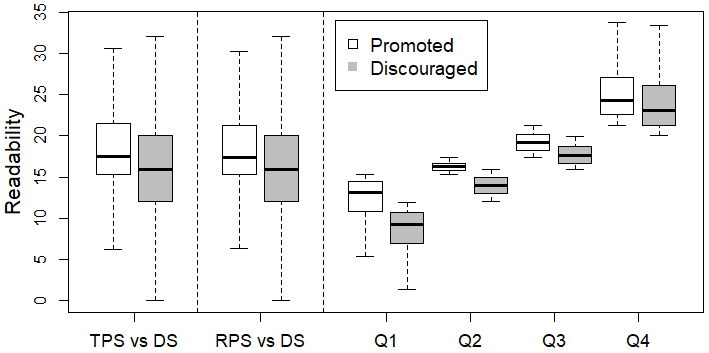}\label{fig:JavascriptTextReadabilityBox(c2)}}
	\hspace{2mm}\subfloat[Python questions with no code snippet]
	{\includegraphics[width=1.68in]{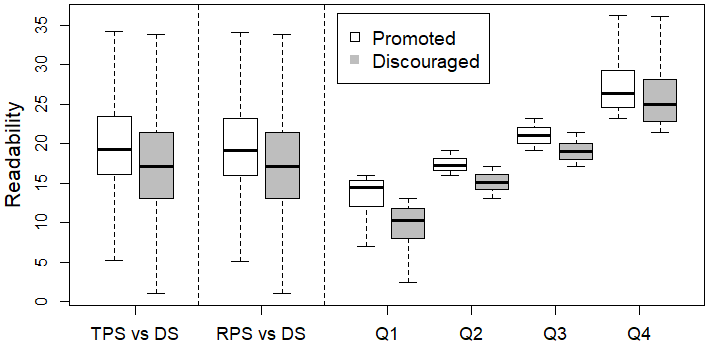}\label{fig:PythonTextReadabilityBox(d1)}}
	\hspace{2mm}\subfloat[Python questions with code snippet]
	{\includegraphics[width=1.68in]{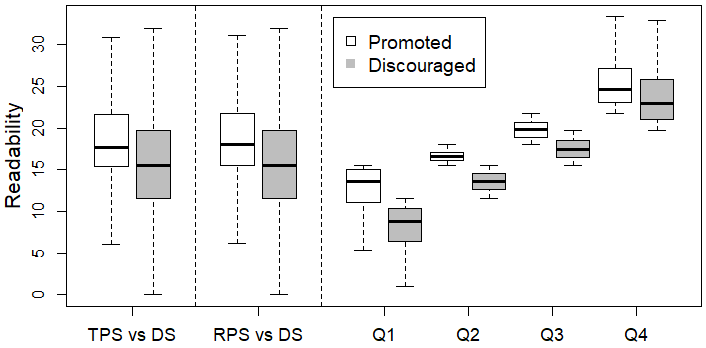}\label{fig:PythonTextReadabilityBox(d2)}}
	
	\caption{Text readability \small{(\textbf{TPS} = Total Promoted Sample, \textbf{RPS} = Random Promoted Sample, \textbf{DS} = Discouraged Sample, \textbf{Q1} = First quartile, \textbf{Q2} = Second quartile, \textbf{Q3} = Third quartile and \textbf{Q4} = Fourth quartile).}}
	\label{fig:text-readability-box-plot}
	
\end{figure*}

\subsection{Readability}
\label{secRead}

Readability is the ease with which a reader can understand a written text \cite{readabilitywiki}. SO questions generally contain two types of items - regular texts and code snippets. We thus attempt to capture the readability of both texts and code using two separate metrics: \emph{Text Readability} and \emph{Code Readability}. We extract code snippets and texts (body + title) by parsing the HTML contents of each question. 
We then apply the readability metrics to them as follows.

%
%

\subsubsection {Text Readability (TR)} 
The readability of texts depends on the complexity of their vocabulary, syntax, and presentation style, such as font size, line height, and line length \cite{readabilitywiki}. We compute the text readability separately for the questions having (1) textual description only and (2) textual description and code snippets.
To assess the text readability, we compute a popular readability index, namely \emph{Readability Index {(RIX)}} \cite{LAR-Anderson-1983}. This readability index represents the comprehension difficulty of English passages based on US grade levels.
First, we calculate the readability of each text and then normalize the score between $0$ and $100$. A low score indicates high readability, whereas a high score indicates poor readability of the question texts.

\textbf{RQ\textsubscript{1}(b): Do the promoted questions have higher text readability than that of the discouraged questions?} First, we select only 25\% of questions that do not have any code snippets in their body. As shown in Fig. \ref{fig:text-readability-box-plot}, we find a significant difference in text readability between promoted and discouraged questions, where promoted questions have higher text readability scores than those of discouraged questions. That is, promoted questions are less readable than that of the discouraged questions. In our dataset, the number of promoted questions is higher than the discouraged questions. We thus undersampled the promoted questions to balance the dataset. Then we compare the readability of discouraged samples with that of the promoted samples. For all the programming languages, we get significant \emph{p-values} from the \emph{Mann-Whitney-Wilcoxon} statistical test and small \emph{effect-size} from \emph{Cliff's delta} test. For Java, we get \emph{p-value $\approx$ 0 \textless 0.05} and \emph{Cliff's d = 0.20 (small)}. We also get similar results for the other three languages - C\#, JavaScript and Python.
Next, we divide the readability scores into four quartiles and compare them using box plots.  We find statically significant differences in text readability (i.e., \emph{p-values \textless 0.05)} for all the quartiles between promoted and discouraged questions. However, the \emph{effect-sizes} differ from small to large.
Second, we measure the readability of the rest 75\% questions with code snippets and textual descriptions. However, we only consider their texts to measure readability. In this case, we also find a significant difference in readability between promoted and discouraged questions.

We further investigate why the promoted questions have lower readability. We find that promoted questions often contain program elements (e.g., API classes, function names) that are considered complex words. Unfortunately, the existing readability metrics are not well designed for handling the code or program elements within the texts. However, we attempt to remove the program elements from the textual description, but it poses several challenges. First, we find that many users do not use appropriate HTML tags (e.g., \texttt{<code> ....</code>}) to format their code elements, which makes it difficult to identify them using tags. Second, regular expressions are not sufficient either to detect these elements since different programming languages have different notations for code elements (e.g., variable names). We thus did not remove the code elements from the texts.

We also experiment with other readability metrics (e.g., Flesch Kincaid Score, Gunning Fog Index). However, they were not able to differentiate between promoted and discouraged questions. Interestingly, the readability scores from the RIX can differentiate between the two types of questions. That is, a combination of multiple readability metrics does not affect our findings while increasing the computational complexity. We thus utilize only RIX to determine the readability of texts.

	
	

\subsubsection{Code Readability (CR)}
\label{codeRead}
Code readability is a property that determines how easily a given code snippet can be read and understood by the developers who did not author the code \cite{ASM-Posnett-2011}. Code readability is strongly related to software maintainability and quality. 
Thus, the underlying idea is - the more readable the code is, the easier it is to reuse and maintain in the long run. Since poor readability of code costs development time and effort, we analyze the readability of the code snippets added to the questions. In particular, we attempt to find whether there is any noticeable difference in the code readability between promoted and discouraged questions.


	

To compute the code readability, we use the tool developed by \citet{LAM-Buse-2010}. Their readability model is trained on human perception of readability or understandability. The model uses several textual features from the code that are likely to affect the humans’ perception of readability and predicts a readability score on a scale from zero to one. One indicates that the source code is highly readable, whereas zero indicates poor readability.

\textbf{RQ\textsubscript{1}(c): Do the code examples of promoted questions have a higher readability score than those of discouraged questions?} 
According to our analysis, the readability of the code snippets found in either promoted or discouraged questions of SO is very poor.
For example, a significant fraction (34\%) of the code snippets have a readability score of zero. Only about 14\% of code snippets have medium or above (i.e., score $\geq 0.5$) readability scores. Although the \emph{p-values} from \emph{Mann-Whitney-Wilcoxon} statistical test are significant (i.e., \emph{p-values} \textless 0.05), the \emph{d-values} from \emph{Cliff's Delta} test show that the \emph{effect sizes} are negligible (i.e., \emph{d \textless 0.1}) for all four programming languages.


To get further insight, we randomly selected 200 code snippets (100 promoted + 100 discouraged) from all four programming languages and spent about ten hours analyzing them. One of the authors, with more than ten years of software development experience, manually investigated the code snippets. Multiple annotators were not involved since we only concentrated on the code snippets' structure, completeness, and noises (e.g., stack traces). However, we first carefully analyzed the tool's specifications \cite{LAM-Buse-2010} to calculate the readability score. Then we check the violations that hurt readability based on those specifications. Our further manual investigation makes the following observations. (1) Most of the code snippets are incomplete, and they do not maintain proper structures, indentations, and naming conventions of identifiers. (2)  Question submitters often include multiple code snippets in a question. When we merge the snippets into a single file, they become large and chaotic. Our adopted tool might not be well suited for such synthesized code snippets. (3) Many of the code snippets are noisy. Programmers often include non-code elements (e.g., stack traces) within the code snippet. Such noise might have hurt their readability.


\subsection{Quality of Explanation for the Code Snippets}
\label{explanation-quality}

In SO questions, users often explain their code snippets with associated prose written in regular texts. \citet{AAO-Ercan-2015} suggest that large and unexplained code snippets make a question difficult to understand, which might impact the time required to obtain an appropriate answer. An explanation that properly explains the code might help the users to answer the question. Thus, the code explanation quality might be considered a metric of question quality. We measure the explanation quality in two ways as follows.

\subsubsection{Text-Code Ratio (TCR)}

The metric determines how extensively the code has been explained by the associated prose in a question. To examine how comprehensive the code explanation is, we extract the code snippets and texts from the body of a question. Then we calculate their lengths in terms of the number of ASCII characters they used. Finally, we measure the text-code ratio as follows.

\smallskip
$TCR = \frac{{LENGTH}(code)}{{LENGTH}(texts)}$
\smallskip

\begin{figure}
	\centering
	
	\pgfplotstableread{
		1	44		35
		2	56		65
		3	44		35
		4	56		65
		
	}\datatable

        \resizebox{2.5in}{!}{%
	\begin{tikzpicture}
	\begin{axis}[
	xtick=data,
	xticklabels={AS$(0 < Ratio \leq 1)$, AS$(Ratio > 1)$, RS$(0 < Ratio \leq 1)$, RS$(Ratio > 1)$
	},
	enlarge y limits=false,
	enlarge x limits=0.2,
	ymin=0,ymax=75,
	ybar,
	bar width=0.8cm,
	width=12cm,
	ylabel={Questions (\%)},
	height = 6cm,
	ymajorgrids=true,
	xticklabel style={font=\footnotesize, /pgf/number format/fixed},	
	major x tick style = {opacity=0},
	minor x tick num = 1,
	ytick={0,20,...,60},
	yticklabels={0\%,20\%,40\%,60\%},    
	minor tick length=1ex,
	legend style={
		font=\footnotesize,
		cells={anchor=west},
		legend columns=1,
		at={(0.15,0.96)},
		anchor=north,
		text width=2cm,
		minimum height=0.4cm,
		/tikz/every even column/.append style={column sep=0.2cm}
	}
	]
	
	\addplot[draw=black!100, fill=black!0] table[x index=0,y index=1] \datatable;
	\addplot[draw=black!100, fill=black!20] table[x index=0,y index=2] \datatable;
	
	\legend	{Promoted,
		Discouraged			
	}
	\end{axis}
	
	\end{tikzpicture}
         }
	\caption{Text-code ratio \small{(\textbf{AS}=All Samples, \textbf{RS}=Random Sampled).}}
	\label{fig:text-code-ratio}
	
\end{figure}
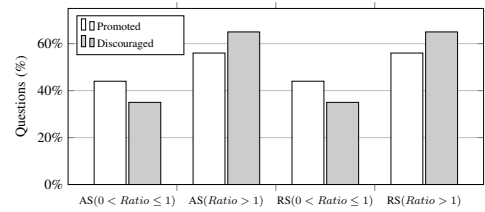

\begin{table}[]
	
	\captionsetup{justification=centering, labelsep=newline}
	\centering
	\caption{Summary of the Text-Code Ratio (TCR) \small{(\textbf{PM}: Promoted, \textbf{DC}: Discouraged)}}
	\label{table:text-code-ratio}
	\resizebox{3.4in}{!}{%
		\begin{tabular}{l|c|c|c|c|c|c|c|c}
			\hline
			\multirow{2}{*}{\textbf{\hspace{14mm}Ratio}}& \multicolumn{2}{c|}{\textbf{C\#}} & \multicolumn{2}{c|}{\textbf{Java}} & \multicolumn{2}{c|}{\textbf{JavaScript}} & \multicolumn{2}{c}{\textbf{Python}} \\ \cline{2-9}
			& PM       & DC      & PM       & DC       & PM          & DC          & PM        & DC        \\ \hline \hline
			Ratio \textless{}= 1 (in percent) & 46             & 36               & 39             & 28                & 44                & 35                   & 51              & 44                 \\ \hline
			Ratio \textgreater 1 (in percent) & 54             & 64               & 61             & 72                & 56                & 65                   & 49              & 56                 \\ \hline
		\end{tabular}
	}
\end{table}

\textbf{RQ\textsubscript{1}(d): Do promoted questions contain more explanation for their code than that of the discouraged questions?} We divide TCR into two categories.
When the length of the texts is larger or equal to the code, these questions belong to the first category.  The second category includes the questions which have a greater code length than that of the texts. From Fig. \ref{fig:text-code-ratio}, we see that the promoted questions have more code explanations than the discouraged questions. \citet{ABO-Squire-2014} also suggest that high-quality questions have a more detailed explanation of code snippets than that of low-quality questions. As shown in Table \ref{table:text-code-ratio}, all the promoted questions of the first category have a higher code explanation ratio than that of discouraged questions from the four programming languages. For example, promoted questions of C\#, Java, JavaScript, and Python have 10\%, 11\%, 9\%, and 7\% higher text-code ratios, respectively, than that of discouraged questions.
We also determine whether the difference in text-code ratio between promoted and discouraged questions is statistically significant. We use the \emph{Mann-Whitney-Wilcoxon} statistical test and find significant p-value \emph{(p-value $\approx$ 0)} with a \emph{small} effect size. 

	

\subsubsection{Text-Code Correlation (TCC)}

The previous section measures the comprehensiveness of the code explanation. In this section, we determine the appropriateness of the code explanation by measuring the correlation between code snippets and code explanations within a question. \citet{AAO-Ercan-2015} evaluate the quality of the explanation of a code snippet in terms of human readability and suggest that a high textual overlap between code snippets and code explanation is strongly correlated to the quality of a question. We thus hypothesize that if there exists token overlap, then the textual description is possibly talking about the code snippet. Therefore, we investigate the content overlapping between code explanations and code snippets using ROUGE. In particular, we use code snippets as a reference summary and code explanations as a system summary. Then we calculate recall. The resulting ROUGE scores (i.e., recall) indicate how appropriately the code explanation (system summary) explains the target code (reference summary).

\begin{figure}
	\centering
	
	
	
	\subfloat[All programming languages]
	{\includegraphics[width=1.7in]{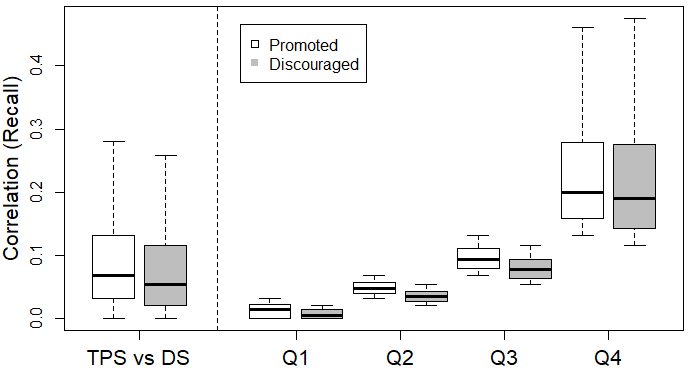}\label{fig:all-rouge-box-plot(e)}}
	\hspace{2mm}\subfloat[All programming languages (After Random Sampling)]
	{\includegraphics[width=1.7in]{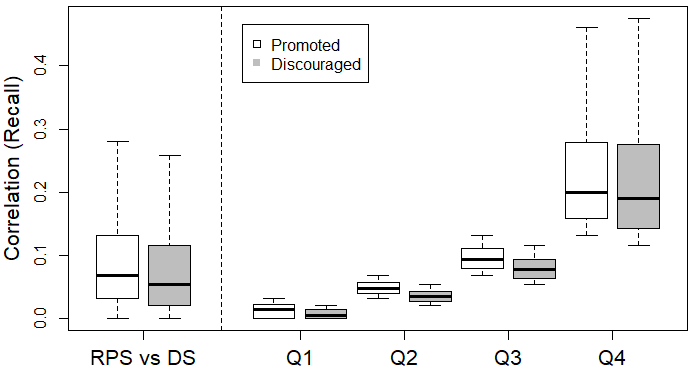}\label{fig:random-rouge-box-plot(f)}}
	
	\caption{Summary of text-code correlation measurement.} 
	
	\label{fig:rough-correlation-summary}
	
\end{figure}

\textbf{RQ\textsubscript{1}(e): Do the promoted questions have a higher correlation between their code snippets and code explanations than those of the discouraged questions?} According to our analysis, the correlation between code snippets and code explanations of promoted questions is stronger than that of the discouraged questions. 
Fig. \ref{fig:rough-correlation-summary} summarizes the results as box plots. We see that all the promoted questions have a higher correlation between explanation and code with significant p-values and small to large effect size.

We use \emph{Mann-Whitney-Wilcoxon} statistical test and \emph{Cliff's delta} test to compute the statistical difference and effect size, respectively. The results of the tests show that the first, second, and third quartiles have a statistically significant difference (i.e., p-values $< 0.05$) with a high effect size. In contrast, the fourth quartile shows a small or negligible effect size.

%
%
%


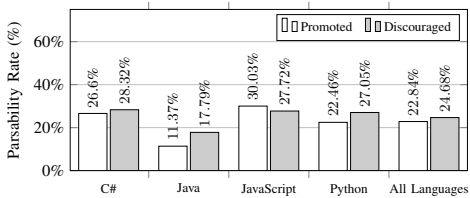
\begin{figure}
	\centering
	
	\pgfplotstableread{
		1	26.60		28.32
		2	11.37		17.79
		3	30.03		27.72
		4	22.46		27.05
		5	22.84		24.68
		
	}\datatable
	\resizebox{2.5in}{!}{%
	\begin{tikzpicture}
	\begin{axis}[
	xtick=data,
	xticklabels={C\#, Java, JavaScript, Python, All Languages
	},
	enlarge y limits=false,
	enlarge x limits=0.12,
	ymin=0,ymax=75,
	ybar,
	bar width=0.6cm,
	width=10cm,
	ylabel={Parsability Rate (\%)},
	ytick={0,20,...,60},
	yticklabels={0\%,20\%,40\%,60\%},
	height = 5cm,
	ymajorgrids=true,
	xticklabel style={font=\footnotesize, /pgf/number format/fixed},	
	major x tick style = {opacity=0},
	minor x tick num = 1,    
	minor tick length=1ex,
	legend style={
		font=\footnotesize,
		cells={anchor=west},
		legend columns=-1,
		at={(0.75,0.97)},
		anchor=north,
		/tikz/every even column/.append style={column sep=0.2cm}
	},
	nodes near coords style={rotate=90,  anchor=west, font=\small},
	nodes near coords,
	nodes near coords =\pgfmathprintnumber{\pgfplotspointmeta}\%
	]
	
	\addplot[draw=black!100, fill=black!0] table[x index=0,y index=1] \datatable;
	\addplot[draw=black!100, fill=black!20] table[x index=0,y index=2] \datatable;
	
	\legend	{Promoted,
		Discouraged			
	}
	\end{axis}
	
	\end{tikzpicture}
        }
	\caption{Parsability rates of the programming languages.}
	\label{fig:parsability}
	
\end{figure}

\subsection{Code Reusability (CRUSE)}

Usability of source code is determined based on parsability, compilability, and executability of source code \cite{FQT-Yang-2016}. In this research, we examine only \emph{parsability} as a part of the reusability of code snippets found in SO questions. Our goal is to measure the parsability rates of the code snippets of four programming languages (C\#, Java, JavaScript, and Python). We then contrast the parsability rate of promoted questions with the discouraged questions.
For parsing C\# code snippets, we use a tool called \emph{Roslyn} by Microsoft. Roslyn provides rich APIs (e.g., CSharpSyntaxTree.ParseText) for parsing the abstract syntax tree from the C\# code. To parse Java code snippets, we used \emph{JavaParser}\footnotemark[1]\footnotetext[1]{http://javaparser.org}. \emph{Esprima}\footnotemark[2]\footnotetext[2]{http://esprima.org} was used to parse JavaScript code examples. Python's built-in AST module was used to parse the Python code snippets.

\textbf{RQ\textsubscript{1}(f): Are code snippets from promoted questions easier to parse than those of the discouraged questions?} As shown in Fig. \ref{fig:parsability}, the parsability rate is higher for the code examples from the discouraged questions than that of the promoted questions. This observation is true for three out of four programming languages - C\#, Java, and Python. 
We find that the overall parsability rate of code examples from promoted questions is about 23\%, whereas such ratio is  25\% for the code snippets of discouraged questions. We further attempt to find why code examples of the discouraged questions have higher parsing rates through manual analysis. According to our investigation, users (especially novice users) at SO often submit questions that contain unnecessarily long code without proper analysis. We believe that such a long code places an extra burden on the other users during question answering. As a result, such questions often receive downvotes and fail to get appropriate answers even though their code is parsable.



	
	

\subsection{Code Understandability (CUA)}

We estimated the readability of the code snippets using an existing tool \cite{LAM-Buse-2010}. However, the perceived readability is something different from the actual understandability of the code. For example, a developer could find a piece of code readable but still challenging to understand what the code does \cite{AAC-Scalabrino-2017}. \citet{AAC-Scalabrino-2017} report that code understandability often depends on the number of APIs used in a program. The more APIs are used, the harder would be the code to comprehend. Thus, we count the number of APIs in the code snippets as a proxy of code understandability. We use regular expressions to find the APIs used within the code snippets.

\textbf{RQ\textsubscript{1}(g): Are code snippets from promoted questions more understandable than those of discouraged questions?} According to our analysis, the difference in the number of APIs usage between promoted and discouraged questions is not statistically significant.

\begin{figure}
	\centering
	
	
	
	\subfloat[All programming languages]
	{\includegraphics[width=1.7in]{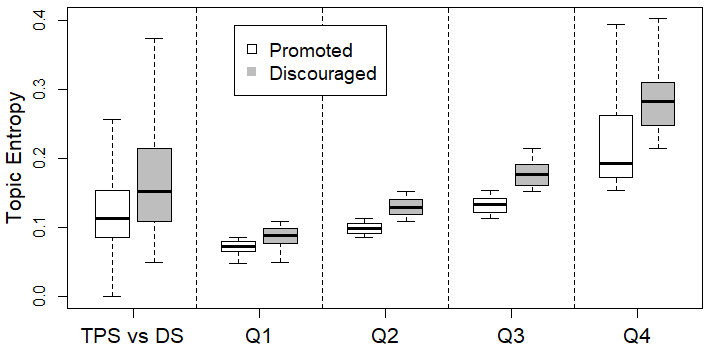}\label{fig:all-te-box-plot(e)}}
	\hspace{2mm}\subfloat[All programming languages (After Random Sampling)]
	{\includegraphics[width=1.7in]{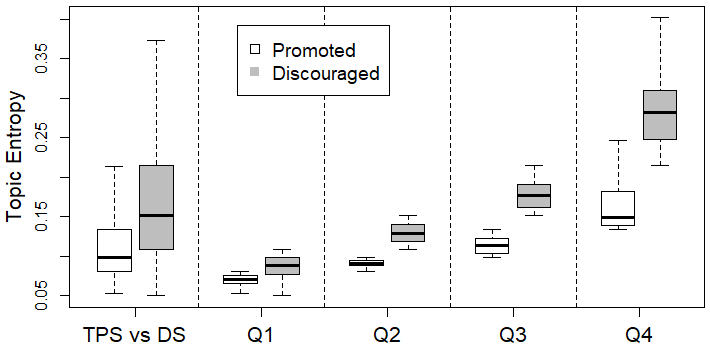}\label{fig:random-te-box-plot(f)}}
	
	\caption{Summary of topic entropy.} 
	
	\label{fig:topic-entropy-summary}
\end{figure}

\subsection{Topic Entropy (TE)}

Programming questions discussing only common topics (e.g., java) are likely to be imprecise or ambiguous. Precise topics could help the users locate the questions of their interest. We study the uncertainty or ambiguity of question topics and attempt to determine if discouraged questions use more common topics than those of promoted questions. In SO, tags capture the topics with which a question is associated \cite{AES-Zhang-2019}. Each question can have at most five tags and must have at least one tag.
In information theory, entropy is considered a measure of uncertainty of a random variable that takes up different values \cite{AII-Rahman-2015}. To measure the topic entropy associated with each question, we first calculate the probability of each topic (i.e., tag). Consider the topic probability of a particular topic $i$ is  $P_i$.

\begin{equation}
	P_i = \frac{F_i}{\sum_{i=1}^{N}F_i}
\end{equation} where $F_i$ is the frequency count of the $i^{th}$ topic and $N$ denotes the total number of topics. We then determine topic entropy for each question as follows.

\begin{equation}
    TE = - \frac{1}{n}\Bigg[\sum_{k=1}^{n} P_{k} \times log(P_{k})\Bigg]
\end{equation} $n$ denotes the total number of topics (i.e., tags) of a question. Here, we calculate the average entropy (dividing the entropy by $n$) since the number of tags differs from question to question. Then we normalize them. If a question of SO uses a set of fairly common topics, the entropy becomes higher. According to information theory, a higher entropy value indicates high ambiguity or uncertainty of question topics and vice versa.

\textbf{RQ\textsubscript{1}(h): Do the promoted questions have less ambiguous topics?} Fig. \ref{fig:topic-entropy-summary} shows the box plots of topic entropy. We see that the topic entropy of promoted questions is lower than that of the discouraged questions for all programming language. In other words, promoted questions have precise topics, whereas the topics of the discouraged questions are ambiguous.
We further test whether the differences in topic entropy are statistically significant for each language. We use the \emph{Mann-Whitney-Wilcoxon} test and find significant p-values (\emph{p-value $\approx$ 0 \textless 0.05}) for all the test cases. Our quartile analysis using \emph{Cliff's delta} test suggests that the differences are significant with a large effect size for all four quartiles. When we take the entire dataset, the difference is also statistically significant with medium or small effect sizes. 


\begin{figure}
	\centering
		
	\subfloat[All programming languages]
	{\includegraphics[width=1.7in]{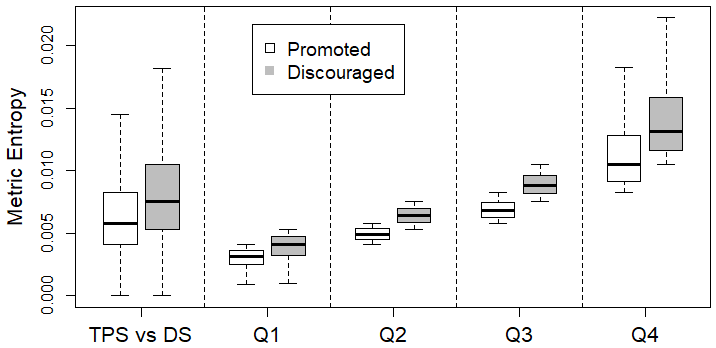}\label{fig:all-me-box-plot(e)}}
	\hspace{2mm}\subfloat[All programming laguages (After Random Sampling)]
	{\includegraphics[width=1.7in]{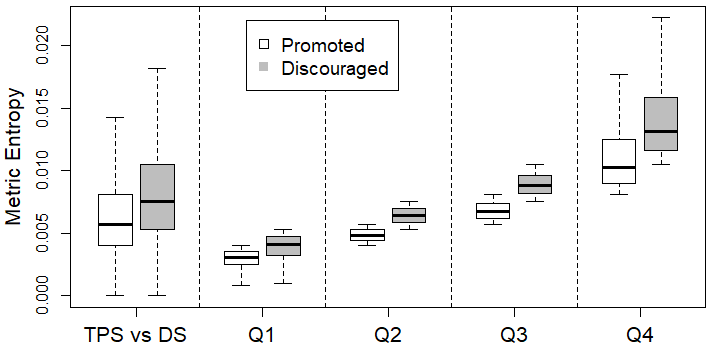}\label{fig:random-me-box-plot(f)}}
	
	\caption{Summary of metric entropy.} 
	
	\label{fig:metric-entropy-summary}
\end{figure}

\subsection{Metric Entropy (ME)}

We use metric entropy to contrast between promoted and discouraged questions. Metric entropy is the Shannon entropy \cite{EOI-Thomas-1991} divided by the length of the text. It represents the randomness of the information contained in a question text \cite{UAC-Ponzanelli-2014}. A higher metric entropy indicates that the questions frequently used common words in their texts and vice versa.

\textbf{RQ\textsubscript{1}(i): Do the promoted questions have used more unusual terms?} Fig. \ref{fig:metric-entropy-summary} shows the box plots of metric entropy. We see that promoted questions have lower metric entropy than that of discourage questions for all the programming languages. That is, promoted questions use more unusual terms, whereas the terms of the discouraged questions are common. We then use \emph{Mann-Whitney-Wilcoxon} test and find significant p-values (i.e., \emph{p $\approx$ 0 \textless 0.05}) for the entire dataset.  However, the effect size is small. Our quartile analysis using \emph{Cliff's delta} test suggests that the differences are significant with a medium or large effect size for all four quartiles.


\begin{table}[!ht]
	\centering
	\captionsetup{justification=centering, labelsep=newline}
	\caption{Details of Sentiment Polarity}
	\label{table:sentiment-summary}
	\resizebox{3.4in}{!}{%
	\begin{tabular}{l|l|c|c|c|c} \hline
		
		\multirow{2}{*}{\textbf{Language}}  & \multirow{2}{*}{\textbf{Category}} & \multicolumn{4}{c}{\textbf{Sentiment Polarity (in-percent)}} \\ \cline{3-6}
		                              &                  &\textbf{Positive} & \textbf{Negative} & \textbf{Mixed} & \textbf{Neutral}  \\ \hline \hline
		
		\multirow{2}{*}{Java}        & Promoted         &  23.60   &  17.26    &  15.46  &  43.68  \\ \cline{2-6}
		                             & Discouraged      &  20.95   &  16.25   &  11.05 &  51.74  \\ \hline
		
		\multirow{2}{*}{C\#}         & Promoted         &  24.03    &  16.78   &  15.53 &  43.66  \\ \cline{2-6}
		                             & Discouraged      &  22.00    &  15.05   &  10.00 &  52.94  \\ \hline
		
		\multirow{2}{*}{Javascript}  & Promoted         &  26.53    &  14.71    &  14.75 &  44.00  \\ \cline{2-6}
		                             & Discouraged      &  24.52    &  12.71    &  9.94 &  52.81  \\ \hline
		
		\multirow{2}{*}{Python}      & Promoted         &  23.48   &  17.47    &  15.77  &  43.29  \\ \cline{2-6}
		                             & Discouraged      &  21.43   &  16.70    &  11.33 &  50.53  \\ \hline
		
	\end{tabular}
}
\vspace{-2mm}
\end{table}

\subsection{Sentiment Polarity (SP)}

Software developers often interact with each other in various community-based Q\&A sites (e.g., SO) and development environments (e.g., Github). \citet{MSA-Calefato-2015} argued that social aspects, such as a developer's emotional status, might impact the response time on Q\&A site. That is, the sentiment expressed in the question texts might determine the answering time of the question. We thus consider the sentiment polarity of SO question texts as an objective quality aspect. We used \emph{SentiStrength-SE}, an sentiment analysis tool, to measure the sentiment polarity. SentiStrength-SE is developed by \citet{LAS-Islam-2017} for the software engineering domain. 

\textbf{RQ\textsubscript{1}(j): Does sentiments expressed in SO question influence their subjective assessment?} We measure \emph{positive, negative, mixed} and \emph{neutral} sentiment polarity of the texts extracted from the title and body of a question. Table \ref{table:sentiment-summary} shows the summary of sentiment measures of the SO questions. We see that about half of the contents from both promoted and discouraged questions are neither positive nor negative. We also see that texts of promoted questions have a higher percentage of positive sentiment and a lower percentage of neutral sentiment polarity than those of discouraged questions. However, none of the differences is statistically significant.

\begin{table}[]
	\centering
	\begin{minipage}{0.22\textwidth}
	\centering
	\captionsetup{justification=centering, labelsep=newline}
	\caption{Feature Ranks (Information Gain)}
	\label{table:info-gain}
	\resizebox{1.5in}{!}{%
	\begin{tabular}{c|l|c}
		\hline
		\textbf{Rank} & \multicolumn{1}{c|}{\textbf{Feature}} & \textbf{Score} \\ \hline \hline
		1             & TE                         & 0.6931         \\ \hline
		2             & ME                        & 0.5328         \\ \hline
		3             & TCR                       & 0.3156         \\ \hline
		4             & CR                      & 0.2803         \\ \hline
		5             & TR                      & 0.2486         \\ \hline
		6             & TCC                 & 0.0167         \\ \hline
		7             & TQ                         & 0.0044         \\ \hline
		8             & SP                    & 0.0040         \\ \hline
		9             & CUA                & 0.0021         \\ \hline
		10            & CRUSE                      & 0.0017         \\ \hline
	\end{tabular}
	}
	\end{minipage}
	\begin{minipage}{0.22\textwidth}
		\centering
		\captionsetup{justification=centering, labelsep=newline}
		\caption{Feature Analysis using Decision Trees}
		\label{table:single-feature-accuracy}
		\resizebox{1.65in}{!}{%
		\begin{tabular}{c|l|c}
			\hline
			\textbf{Rank} & \multicolumn{1}{c|}{\textbf{Feature}} & \textbf{Accuracy} \\ \hline \hline
			1             & TE                         & 66.36\%         \\ \hline
			2             & TR                      & 63.44\%         \\ \hline
			3             & ME                        & 60.49\%         \\ \hline
			4             & TCR                       & 56.25\%         \\ \hline
			5             & TCC                 & 54.11\%         \\ \hline
			6             & SP                    & 53.95\%         \\ \hline
			7             & CUA                & 53.01\%         \\ \hline
			8             & TQ                         & 52.86\%         \\ \hline
			9             & CR                      & 52.18\%         \\ \hline
			10            & CRUSE                      & 52.17\%         \\ \hline
		\end{tabular}
		}
	\end{minipage}
\end{table}

\section{Construction of Models to Classify Questions based on their Quality (RQ$_2$)}
\label{prediction}

In this section, we first discuss how we ranked the features found from the section \ref{comparativeStudy}. Then we describe our selected machine learning models with their settings. Next, we evaluate the performances of our models. Finally, we compare our models' performances with the baseline models.

\subsection{Ranking Features}
\label{featureRanking}



In section \ref{comparativeStudy}, we showed several attributes (e.g., topic entropy) whose values differ for promoted and discouraged questions. Such findings indicate that these features might be the key estimators in predicting whether a question will be up-voted or down-voted. However, we do not know yet which attributes are more important than others in differentiating up-voted (i.e., promoted) and down-voted (i.e., discouraged) questions. Therefore, a ranking of the attributes might help to select the top $N$ features in predicting questions. We use two popular measures to rank the features as follows.

\textbf{Information Gain.}
In information theory, the Information Gain of a random variable is the change in information entropy between a prior state and a state that takes some information \cite{TUT-Saha-2013}. Therefore, the Information Gain of a feature predicting whether the questions will be promoted or discouraged is as follows.

\begin{equation}
	Info\hspace{0.5mm}Gain(C, a{_i}) = H(C) - H(C|a{_i})
\end{equation} 
where $C$ represents a particular class (i.e., promoted or discouraged), $a{_i}$ denotes the attribute, and $H$ denotes information entropy. As shown in Table \ref{table:info-gain}, topic entropy has the highest information gain. That is, topic entropy might discriminate the questions with the maximum accuracy. Metric entropy, text-code ratio, code readability, and text readability have the second, third, fourth and fifth highest information gain respectively. However, gain ratio of the remaining attributes is either small or negligible. 

	
	
	

\textbf{Analysis with Decision Trees.} 
In this study, we not only attempt to predict the quality of a question during question submission time but also to understand which attributes influence the quality of a question. We thus attempt to see how each attribute can classify the promoted and discouraged questions using Decision Trees. Table \ref{table:single-feature-accuracy} shows the classification accuracy of each of the features. We see that topic entropy can classify the questions with overall 66.36\% accuracy. Text readability, metric entropy, and text-code ratio have the second, third, and fourth highest discrimination power, respectively. As opposed to information gain, we find that text-code correlation has a higher classification accuracy than code readability. However, four of the top five attributes are common in both cases. Thus, we select these four features (e.g., topic entropy, text readability, metric entropy, code-text ratio) as an optimal feature set and analyze the models' performance using all the features and the top four separately.

\subsection{Classification Models and Settings}

\textbf{Model Selection.}
According to our comparative study, the relationship between question classes and their corresponding feature values might be complex. 
We thus choose five popular supervised machine learning techniques with different learning strategies. They are -- i) Decision Tree~\cite{quinlan1986induction, cart} (DT) ii) Random Forest (RF)~\cite{randomForest}, iii) Artificial Neural Network (ANN)~\cite{ann}, iv) K-Nearest Neighbors (KNN)~\cite{knn}, and v) Gaussian Naive Bayes (GNB)~\cite{webb2010naive, rish2001empirical}. In particular, we choose these machine learning algorithms because they are widely used in the relevant studies \citep{TUT-Saha-2013,UAC-Ponzanelli-2014,AII-Rahman-2015, beyer2018automatically}. We thus believe they can build reliable models to classify promoted and discoursed questions. We use \emph{Scikit-learn} \cite{SML-Pedregosa-2011}, one of the popular and widely used tools to implement the techniques.

\textbf{Parameter Tuning.} 
Tuning parameters in classifiers is important because it changes the heuristics determining how they learn \cite{AEA-Calefato-2019}. For example, it controls the number of decision trees to use in RF or the number of clusters in KNN. Models trained with suboptimal parameter settings may underperform as parameter settings depend on the dataset \cite{ASL-Hall-2012}. To select the best model configuration, we use \emph{GridSearchCV}, the cross-validated grid search algorithm of Scikit-learn \cite{SML-Pedregosa-2011}. The GridSearchCV algorithm works by running and evaluating the model performance of all possible combinations of parameters provided to it. 
We choose a number of plausible options for tuning the important parameters of the machine learning models.  GridSearchCV does two tasks in parallel (1) cross-validation and (2) parameter tuning. Finally, it returns the values for a model's parameters that maximize the accuracy of the model. Here, we keep the default values for the optional parameters. We also investigate the efficiency of both training and test split set and choose the parameters thereby to avoid model overfitting.

\begin{table*}[]
	\centering
	\captionsetup{justification=centering, labelsep=newline}
	\caption{Performance of Classifiers (\textbf{H} indicates the highest values)}
	\label{table:performance-summary}
	\resizebox{5.4in}{!}{%
		\begin{tabular}{l|l|l|l|l|l|l|l|l|c}
			\hline
			\multicolumn{1}{c|}{\multirow{2}{*}{\textbf{Classifier}}} & \multicolumn{1}{c|}{\multirow{2}{*}{\textbf{Features}}} & \multicolumn{1}{c|}{\multirow{2}{*}{\textbf{Dataset}}} & \multicolumn{3}{c|}{\textbf{Promoted Question}}  & \multicolumn{3}{c|}{\textbf{Discouraged Question}}  & \textbf{Overall}  \\ \cline{4-10} 
			
			\multicolumn{1}{c|}{}  & \multicolumn{1}{c|}{} & \multicolumn{1}{c|}{} & \multicolumn{1}{c|}{\textbf{Precision}} & \multicolumn{1}{c|}{\textbf{Recall}} & \multicolumn{1}{c|}{\textbf{F1-Score}} & \multicolumn{1}{c|}{\textbf{Precision}} & \multicolumn{1}{c|}{\textbf{Recall}} & \multicolumn{1}{c|}{\textbf{F1-Score}} & \textbf{Accuracy} \\ \hline\hline
			
			\multirow{4}{*}{\textbf{DT}} & \multirow{2}{*}{\textbf{All}} & \textbf{Imbalanced}  & 90.90 & 89.98 & 90.44 & 59.27 & 61.80 & 60.51 \textbf{H}& 84.60 \\ \cline{3-10} 
			&                                                            & \textbf{Balanced}    & 74.56 & 74.17 & 74.37 & 74.30 & 74.70 & 74.50 & 74.30 \\ \cline{2-10} 
			& \multirow{2}{*}{\textbf{Top-4}}                            & \textbf{Imbalanced}  & 91.85 & 91.10 & 91.48 \textbf{H}& 63.51 & 65.73 & 64.60 \textbf{H}& 86.26 \textbf{H} \\ \cline{3-10} 
			&                                                            & \textbf{Balanced}    & 75.37 & 74.90 & 75.14 \textbf{H}& 75.06 & 75.53 & 75.29 \textbf{H}& 75.21 \textbf{H} \\ \hline\hline
			
			\multirow{4}{*}{\textbf{RF}} & \multirow{2}{*}{\textbf{All}} & \textbf{Imbalanced}  & 88.39 & 94.03 & 91.12 \textbf{H}& 65.29 & 47.65 & 55.09 & 85.18 \textbf{H} \\ \cline{3-10} 
			&                                                            & \textbf{Balanced}    & 76.98 & 72.09 & 74.46 & 73.76 & 78.44 & 76.03 \textbf{H}& 75.27 \\ \cline{2-10} 
			& \multirow{2}{*}{\textbf{Top-4}}                            & \textbf{Imbalanced}  & 89.01 & 94.45 & 91.45 & 68.22 & 50.54 & 58.06 & 86.07 \\ \cline{3-10} 
			&                                                            & \textbf{Balanced}    & 75.17 & 71.90 & 73.50 & 73.07 & 76.25 & 74.62 & 74.07 \\ \hline\hline

			\multirow{4}{*}{\textbf{ANN}} & \multirow{2}{*}{\textbf{All}} & \textbf{Imbalanced}  & 83.01 & 97.45 & 89.65 & 58.79 & 15.40 & 24.41 & 81.80 \\ \cline{3-10} 
			&                                                             & \textbf{Balanced}    & 73.54 & 79.41 & 76.36 \textbf{H}& 77.62 & 71.42 & 74.39 & 75.42 \textbf{H} \\ \cline{2-10} 
			& \multirow{2}{*}{\textbf{Top-4}}                             & \textbf{Imbalanced}  & 76.42 & 89.80 & 82.57 & 68.61 & 44.60 & 54.06 & 74.73 \\ \cline{3-10} 
			&                                                             & \textbf{Balanced}    & 70.37 & 75.79 & 72.98 & 73.77 & 68.08 & 70.81 & 71.94 \\ \hline\hline
			
			\multirow{4}{*}{\textbf{KNN}} & \multirow{2}{*}{\textbf{All}} & \textbf{Imbalanced}  & 76.11 & 88.77 & 81.95 & 66.33 & 44.27 & 53.10 & 73.93 \\ \cline{3-10} 
			&                                                             & \textbf{Balanced}    & 66.87 & 74.71 & 70.57 & 71.35 & 62.99 & 66.91 & 68.85 \\ \cline{2-10} 
			& \multirow{2}{*}{\textbf{Top-4}}                             & \textbf{Imbalanced}  & 76.14 & 88.74 & 81.96 & 66.34 & 44.37 & 53.17 & 73.95 \\ \cline{3-10} 
			&                                                             & \textbf{Balanced}    & 66.43 & 73.61 & 69.74 & 70.41 & 62.80 & 66.39 & 68.21 \\ \hline\hline
			
	\multirow{4}{*}{\textbf{GNB}} & \multirow{2}{*}{\textbf{All}} & \textbf{Imbalanced}  & 82.52 & 95.19 & 88.40 & 41.51 & 14.49  & 21.48 & 79.79 \\ \cline{3-10} 
			&                                                             & \textbf{Balanced}    & 55.73 & 90.60 & 69.01 & 74.88 & 28.01 & 40.77 & 59.31 \\ \cline{2-10} 
			& \multirow{2}{*}{\textbf{Top-4}}                             & \textbf{Imbalanced}  & 82.59 & 95.97 & 88.78 & 45.36 & 14.20 & 21.63 & 80.37 \\ \cline{3-10} 
			&                                                             & \textbf{Balanced}    & 56.62 & 89.59 & 69.39 & 75.07 & 31.35 & 44.23 & 60.47 \\ \hline\hline
			
		\end{tabular}
	}
\end{table*}

\subsection{Metrics of Success}

We evaluate the performance of our classification models using four popular performance metrics -- (1) \emph{precision}, (2) \emph{recall}, (3) \emph{F1-score}, and (4) \emph{classification accuracy}.

\subsection{Performance Analysis}

We apply 10-fold cross-validation for training and testing each machine learning model. We train and test these models using (1) all the features and (2) the top four features. We then analyze the results separately. Table \ref{table:performance-summary} summarizes the classification results. We marked the highest performance by \textbf{H}.
DT classifies the promoted questions with a maximum of about 91\% precision and 90\% recall when we use all of the features and an imbalanced dataset. The precision is about 60\% for the discouraged questions for the imbalanced dataset. The imbalanced dataset might generate suboptimal classification models. The precision increased from about 60\% to 75\% when we trained the model on the balanced dataset. The overall classification accuracies for the balanced and imbalanced datasets are about 75\% and 85\%, respectively. 
When we choose the top-4 features, the evaluation metrics show almost similar results as above. It indicates that the remaining features have a low effect on the classification models. When we train the models with RF and an imbalanced dataset, we see that the precision of the discouraged questions increases from about 60\% to 66\% but recall decreases by about 14\%. Similar results were obtained when we trained the models with top-4 features. 

Among the remaining three machine learning models, ANN performs reasonably well when we use a balanced dataset, and  KNN performs well while using top-4 features. KNN usually performs better with fewer features, and it is not much dependent on the training phase. The ANN and GNB perform poorly in predicting the discouraged questions when we train them with an imbalanced dataset. GNB classifier makes a very strong assumption on the shape of the data distribution. That is, given the output class, it assumes that any two features are either independent or not. Such an assumption might affect the model's performance. GNB also suffers from imbalanced (resulting in skewed probabilities) classes and continuous features \cite{TTP-Rennie-2003}.

After analyzing the results, it could be concluded that tree-based classifiers (e.g., DT) can generate a better model to classify the questions based on our dataset and features. On the other hand, probabilistic models like GNB could not generate optimal models, especially using an imbalanced dataset.

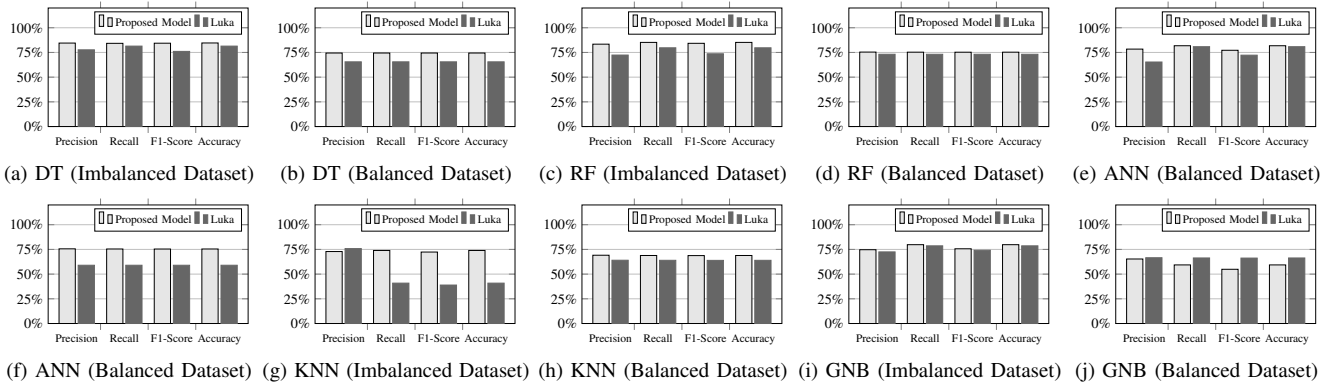
\begin{figure*}
	\centering
	\pgfplotstableread{
		1	84.50  77.66     74.43   65.51     83.39  72.38      75.37  73.18     78.39	 65.34		75.58  58.99  72.85	 75.83   69.11  64.06   74.70	72.49	65.30	66.59
		2	84.23  81.43     74.43   65.51     85.18  79.71      75.27  73.16	  81.80	 80.83		75.42  58.99  73.93  40.88   68.85  63.98   79.79	78.63	59.31	66.36
		3	84.36  76.07     74.43   65.51     84.25  73.66      75.24  73.16 	  77.20	 72.26		75.38  58.98  72.33  38.94   68.74  63.91   75.63	74.16	54.89	66.24
		4	84.60  81.43     74.43   65.51     85.18  79.71      75.27  73.16	  81.80	 80.83		75.42  58.99  73.93  40.88   68.85  63.97   79.79	78.63	59.31	66.35
	}\datatable
	
	\subfloat[DT (Imbalanced Dataset)]{
            \resizebox{1.3in}{!}{%
		\begin{tikzpicture}
		\begin{axis}[
		xtick=data,
		xticklabels={Precision, Recall, F1-Score, Accuracy
		},
		enlarge y limits=false,
		enlarge x limits=0.2,
		ymin=0,ymax=120,
		ybar,
		ytick={0,25,...,100},
		yticklabels={0\%,25\%,50\%,75\%,100\%},
		bar width=0.4cm,
		width=6.5cm,
		height = 4.5cm,
		ymajorgrids=true,
		xticklabel style={font=\footnotesize, /pgf/number format/fixed},
		major x tick style = {opacity=0},
		minor x tick num = 1,    
		minor tick length=1ex,
		legend style={at={(0.6,0.97)},
			font=\footnotesize,
			anchor=north,legend columns=-1},
		]
		\addplot[draw=black!100, fill=black!10] table[x index=0,y index=1] \datatable;
		\addplot[draw=black!60, fill=black!60] table[x index=0,y index=2] \datatable;
		\legend	{Proposed Model,
			Luka
		}
		\end{axis}
		\end{tikzpicture}
            }
		\label{fig:comparision-dt-unbalanced}
	}
	\subfloat[DT (Balanced Dataset)]{
            \resizebox{1.3in}{!}{%
		\begin{tikzpicture}
		\begin{axis}[
		xtick=data,
		xticklabels={Precision, Recall, F1-Score, Accuracy
		},
		enlarge y limits=false,
		enlarge x limits=0.2,
		ymin=0,ymax=120,
		ybar,
		ytick={0,25,...,100},
		yticklabels={0\%,25\%,50\%,75\%,100\%},
		bar width=0.4cm,
		width=6.5cm,
		height = 4.5cm,
		ymajorgrids=true,
		xticklabel style={font=\footnotesize, /pgf/number format/fixed},	
		major x tick style = {opacity=0},
		minor x tick num = 1,    
		minor tick length=1ex,
		legend style={at={(0.6,0.97)},
			font=\footnotesize,
			anchor=north,legend columns=-1},
		]
		\addplot[draw=black!100, fill=black!10] table[x index=0,y index=3] \datatable;
		\addplot[draw=black!60, fill=black!60] table[x index=0,y index=4] \datatable;
		\legend	{Proposed Model,
			Luka
		}
		\end{axis}
		\end{tikzpicture}
            } 
		\label{fig:comparision-dt-balanced}
	}
	\subfloat[RF (Imbalanced Dataset)]{
		\resizebox{1.3in}{!}{%
		\begin{tikzpicture}
		\begin{axis}[
		xtick=data,
		xticklabels={Precision, Recall, F1-Score, Accuracy
		},
		enlarge y limits=false,
		enlarge x limits=0.2,
		ymin=0,ymax=120,
		ybar,
		ytick={0,25,...,100},
		yticklabels={0\%,25\%,50\%,75\%,100\%},
		bar width=0.4cm,
		width=6.5cm,
		height = 4.5cm,
		ymajorgrids=true,
		xticklabel style={font=\footnotesize, /pgf/number format/fixed},	
		major x tick style = {opacity=0},
		minor x tick num = 1,    
		minor tick length=1ex,
		legend style={at={(0.6,0.97)},
			font=\footnotesize,
			anchor=north,legend columns=-1},
		]
		\addplot[draw=black!100, fill=black!10] table[x index=0,y index=5] \datatable;
		\addplot[draw=black!60, fill=black!60] table[x index=0,y index=6] \datatable;
		\legend	{Proposed Model,
			Luka
		}
		\end{axis}
		\end{tikzpicture}
	}
		\label{fig:comparision-rf-unbalanced}
	}
	\subfloat[RF (Balanced Dataset)]{
            \resizebox{1.3in}{!}{%
		\begin{tikzpicture}
		\begin{axis}[
		xtick=data,
		xticklabels={Precision, Recall, F1-Score, Accuracy
		},
		enlarge y limits=false,
		enlarge x limits=0.2,
		ymin=0,ymax=120,
		ybar,
		ytick={0,25,...,100},
		yticklabels={0\%,25\%,50\%,75\%,100\%},
		bar width=0.4cm,
		width=6.5cm,
		height = 4.5cm,
		ymajorgrids=true,
		xticklabel style={font=\footnotesize, /pgf/number format/fixed},	
		major x tick style = {opacity=0},
		minor x tick num = 1,    
		minor tick length=1ex,
		legend style={at={(0.6,0.97)},
			font=\footnotesize,
			anchor=north,legend columns=-1},
		]
		\addplot[draw=black!100, fill=black!10] table[x index=0,y index=7] \datatable;
		\addplot[draw=black!60, fill=black!60] table[x index=0,y index=8] \datatable;
		\legend	{Proposed Model,
			Luka
		}
		\end{axis}
		\end{tikzpicture}
            }
		\label{fig:comparision-rf-balanced}
	}
    \subfloat[ANN (Balanced Dataset)]{
            \resizebox{1.3in}{!}{%
    	\begin{tikzpicture}
    	\begin{axis}[
    	xtick=data,
    	xticklabels={Precision, Recall, F1-Score, Accuracy
    	},
    	enlarge y limits=false,
    	enlarge x limits=0.2,
    	ymin=0,ymax=120,
    	ybar,
    	ytick={0,25,...,100},
    	yticklabels={0\%,25\%,50\%,75\%,100\%},
    	bar width=0.4cm,
    	width=6.5cm,
    	height = 4.5cm,
    	ymajorgrids=true,
    	xticklabel style={font=\footnotesize, /pgf/number format/fixed},	
    	major x tick style = {opacity=0},
    	minor x tick num = 1,    
    	minor tick length=1ex,
    	legend style={at={(0.6,0.97)},
    		font=\footnotesize,
    		anchor=north,legend columns=-1},
    	]
    	\addplot[draw=black!100, fill=black!10] table[x index=0,y index=9] \datatable;
    	\addplot[draw=black!60, fill=black!60] table[x index=0,y index=10] \datatable;
    	\legend	{Proposed Model,
    		Luka
    	}
    	\end{axis}
    	\end{tikzpicture}
            }
    	\label{fig:comparision-ann-unbalanced}
    }

    \vspace{-2mm}
    \subfloat[ANN (Balanced Dataset)]{
            \resizebox{1.3in}{!}{%
    	\begin{tikzpicture}
    	\begin{axis}[
    	xtick=data,
    	xticklabels={Precision, Recall, F1-Score, Accuracy
    	},
    	enlarge y limits=false,
    	enlarge x limits=0.2,
    	ymin=0,ymax=120,
    	ybar,
    	ytick={0,25,...,100},
    	yticklabels={0\%,25\%,50\%,75\%,100\%},
    	bar width=0.4cm,
    	width=6.5cm,
    	height = 4.5cm,
    	ymajorgrids=true,
    	xticklabel style={font=\footnotesize, /pgf/number format/fixed},	
    	major x tick style = {opacity=0},
    	minor x tick num = 1,    
    	minor tick length=1ex,
    	legend style={at={(0.6,0.97)},
    		font=\footnotesize,
    		anchor=north,legend columns=-1},
    	]
    	\addplot[draw=black!100, fill=black!10] table[x index=0,y index=11] \datatable;
    	\addplot[draw=black!60, fill=black!60] table[x index=0,y index=12] \datatable;
    	\legend	{Proposed Model,
    		Luka
    	}
    	\end{axis}
    	\end{tikzpicture}
            }
    	\label{fig:comparision-ann-balanced}
    }
    \subfloat[KNN (Imbalanced Dataset)]{
            \resizebox{1.3in}{!}{%
    	\begin{tikzpicture}
    	\begin{axis}[
    	xtick=data,
    	xticklabels={Precision, Recall, F1-Score, Accuracy
    	},
    	enlarge y limits=false,
    	enlarge x limits=0.2,
    	ymin=0,ymax=120,
    	ybar,
    	ytick={0,25,...,100},
    	yticklabels={0\%,25\%,50\%,75\%,100\%},
    	bar width=0.4cm,
    	width=6.5cm,
    	height = 4.5cm,
    	ymajorgrids=true,
    	xticklabel style={font=\footnotesize, /pgf/number format/fixed},	
    	major x tick style = {opacity=0},
    	minor x tick num = 1,    
    	minor tick length=1ex,
    	legend style={at={(0.6,0.97)},
    		font=\footnotesize,
    		anchor=north,legend columns=-1},
    	]
    	\addplot[draw=black!100, fill=black!10] table[x index=0,y index=13] \datatable;
    	\addplot[draw=black!60, fill=black!60] table[x index=0,y index=14] \datatable;
    	\legend	{Proposed Model,
    		Luka
    	}
    	\end{axis}
    	\end{tikzpicture}
            }
    	\label{fig:comparision-knn-unbalanced}
    }
    \subfloat[KNN (Balanced Dataset)]{
            \resizebox{1.3in}{!}{%
    	\begin{tikzpicture}
    	\begin{axis}[
    	xtick=data,
    	xticklabels={Precision, Recall, F1-Score, Accuracy
    	},
    	enlarge y limits=false,
    	enlarge x limits=0.2,
    	ymin=0,ymax=120,
    	ybar,
    	ytick={0,25,...,100},
    	yticklabels={0\%,25\%,50\%,75\%,100\%},
    	bar width=0.4cm,
    	width=6.5cm,
    	height = 4.5cm,
    	ymajorgrids=true,
    	xticklabel style={font=\footnotesize, /pgf/number format/fixed},	
    	major x tick style = {opacity=0},
    	minor x tick num = 1,    
    	minor tick length=1ex,
    	legend style={at={(0.6,0.97)},
    		font=\footnotesize,
    		anchor=north,legend columns=-1},
    	]
    	\addplot[draw=black!100, fill=black!10] table[x index=0,y index=15] \datatable;
    	\addplot[draw=black!60, fill=black!60] table[x index=0,y index=16] \datatable;
    	\legend	{Proposed Model,
    		Luka
    	}
    	\end{axis}
    	\end{tikzpicture}
            }
    	\label{fig:comparision-knn-balanced}
    }
    \subfloat[GNB (Imbalanced Dataset)]{
            \resizebox{1.3in}{!}{%
    	\begin{tikzpicture}
    	\begin{axis}[
    	xtick=data,
    	xticklabels={Precision, Recall, F1-Score, Accuracy
    	},
    	enlarge y limits=false,
    	enlarge x limits=0.2,
    	ymin=0,ymax=120,
    	ybar,
    	ytick={0,25,...,100},
    	yticklabels={0\%,25\%,50\%,75\%,100\%},
    	bar width=0.4cm,
    	width=6.5cm,
    	height = 4.5cm,
    	ymajorgrids=true,
    	xticklabel style={font=\footnotesize, /pgf/number format/fixed},	
    	major x tick style = {opacity=0},
    	minor x tick num = 1,    
    	minor tick length=1ex,
    	legend style={at={(0.6,0.97)},
    		font=\footnotesize,
    		anchor=north,legend columns=-1},
    	]
    	\addplot[draw=black!100, fill=black!10] table[x index=0,y index=17] \datatable;
    	\addplot[draw=black!60, fill=black!60] table[x index=0,y index=18] \datatable;
    	\legend	{Proposed Model,
    		Luka
    	}
    	\end{axis}
    	\end{tikzpicture}
            }
    	\label{fig:comparision-gnb-unbalanced}
    }
    \subfloat[GNB (Balanced Dataset)]{
            \resizebox{1.3in}{!}{%
    	\begin{tikzpicture}
    	\begin{axis}[
    	xtick=data,
    	xticklabels={Precision, Recall, F1-Score, Accuracy
    	},
    	enlarge y limits=false,
    	enlarge x limits=0.2,
    	ymin=0,ymax=120,
    	ybar,
    	ytick={0,25,...,100},
    	yticklabels={0\%,25\%,50\%,75\%,100\%},
    	bar width=0.4cm,
    	width=6.5cm,
    	height = 4.5cm,
    	ymajorgrids=true,
    	xticklabel style={font=\footnotesize, /pgf/number format/fixed},	
    	major x tick style = {opacity=0},
    	minor x tick num = 1,    
    	minor tick length=1ex,
    	legend style={at={(0.6,0.97)},
    		font=\footnotesize,
    		anchor=north,legend columns=-1},
    	]
    	\addplot[draw=black!100, fill=black!10] table[x index=0,y index=19] \datatable;
    	\addplot[draw=black!60, fill=black!60] table[x index=0,y index=20] \datatable;
    	\legend	{Proposed Model,
    		Luka
    	}
    	\end{axis}
    	\end{tikzpicture}
            }
    	\label{fig:comparision-gnb-balanced}
    }

\caption{Performance comparison of our model with baseline model.} 
\label{fig:comparison-with-baseline}

\end{figure*}

\subsection{Performance Comparison with Baseline Model}
\label{comp-study}

To the best of our knowledge, agreement analysis between subjectivity and objectivity on Stack Overflow is still an unmet attempt. However, \citet{UAC-Ponzanelli-2014} attempt to classify the questions based on quality and build a machine learning model to discriminate the bad (\emph{i.e., score} \textless $0$) and good quality (\emph{i.e., score} \textgreater $0$) questions. They identify three sets of metrics that include textual and non-textual features of SO questions. The metrics are SO ($M{_{SO}}$), readability ($M_R$) and user popularity ($M_P$). However, the attributes of the user popularity (e.g., upvotes, downvotes) metric are directly or indirectly associated with the subjective evaluation. In this paper, we investigate objective quality metrics. We thus select the $M{_{SO}}$ and $M_R$ metrics to prepare an equivalence set of attributes.

Fig. \ref{fig:comparison-with-baseline} shows the comparative results of our models and the baseline models by \citet{UAC-Ponzanelli-2014}. We use the same settings of the machine learning models and identical dataset to compare the models' performances.  First, we calculate precision, recall, F1-score, and overall accuracy. Then we compare the values of each of the evaluation metrics between our models and baseline models. When we train our model with decision trees and use an imbalanced dataset, both the F1-score and accuracy are about 85\% (Fig. \ref{fig:comparision-dt-unbalanced}). They are as low as about 76\% and 82\%, when we train the model with the attributes used in the baseline study \cite{UAC-Ponzanelli-2014}. The performance is also significantly higher when we use a balanced dataset. The F1-score and overall accuracy are about 10\% higher than the baseline model (Fig. \ref{fig:comparision-dt-balanced}). We see that the overall performance of the baseline model improves when we train the model with RF. Nonetheless, our models outperform the baseline models. The F1-score is 10\% higher when we use an imbalanced dataset (Fig. \ref{fig:comparision-rf-unbalanced}). 
Our models also outperforms the baseline models when we train them using ANN (Fig. \ref{fig:comparision-ann-unbalanced} 
\& Fig. \ref{fig:comparision-ann-balanced}) and KNN (Fig. \ref{fig:comparision-knn-unbalanced} 
\& Fig. \ref{fig:comparision-knn-balanced}). However, as shown in Fig. \ref{fig:comparision-gnb-balanced}, the model by \citet{UAC-Ponzanelli-2014} outperforms our model only when we train the model with GNB and use a balanced dataset. The underlying data distribution and independent nature of a few of their features might explain the higher performance of their model when using GNB. However, our analysis using five machine learning classifiers with diverse working principles concludes that our features have higher discrimination power to classify promoted and discouraged questions than those of the baseline models.

\section{Findings Summary}
\label{discussion}

Our study exposes several insights that might prevent a question from getting downvotes.

\textbf{Be aware of choosing topics.} Questions discussing only common topics (e.g., Python, Java) could prevent the users from locating the questions of their interest. Users often discourage the use of improper topics (e.g., tags) by downvotes. According to our investigation, discouraged questions use more common topics than promoted ones.

\textbf{Code reusability does not always guarantee high quality.}
Only those code statements should be added that are required to understand reported issues and redundant code should be avoided. Long and redundant code waste the developers' time, which might also increase the chance of getting downvotes. 

\textbf{Lack of code explanation hurts.}
A clear explanation of code examples might influence the question quality.  We also find that the usage of appropriate technical terms is important.  A short and non-technical explanation might mislead the users from understanding the actual problem.

\textbf{Sentiment polarity is not a big concern.}
A few recent studies report that sentiment polarity (e.g., positive, negative) might influence the quality of posts. \citet{MSA-Calefato-2015} find that the sentiment expressed in the answerers' comments might help to estimate answers' quality. According to our analysis, sentiment polarity has less impact on the subjective quality assessment of SO questions.  In fact, about 50\% of the questions have neither positive nor negative sentiments.

\section{Related Work }
\label{relatedWork}

Several aspects of Stack Overflow have been investigated in recent studies due to the growing popularity and importance of this Q\&A site. Most of the previous studies in mining Q\&A sites concentrate on assessing the quality of the posts \cite{UAC-Ponzanelli-2014, ILQ-Ponzanelli-2014, ITT-Xia-2016, FAR-Roy-2018, PTQ-Baltadzhieva-2015, AEA-Calefato-2019}, understanding the way how developers interact among them on Q\&A sites \cite{HDP-Treude-2011}, providing empirical evidence on writing high-quality posts \cite{HTA-Calefato-2018, BRI-Bosu-2013}, the usability of the code snippets attached in the answers \cite{FQT-Yang-2016, GET-Horton-2018}, and the impact of the sentiment polarity on getting answers \cite{MSA-Calefato-2015, TDT-Novielli-2014}. To the best of our knowledge, this is the first attempt to examine the subjective evaluation of SO with objective quality metrics.

\citet{UAC-Ponzanelli-2014,ILQ-Ponzanelli-2014} devise an approach to classify the questions based on their quality. In particular, they identify the low-quality questions to improve the capability of the manual review process of SO. They define three sets of metrics to measure the questions' quality, namely SO ($M{_{SO}}$), readability ($M_R$), and user popularity ($M_P$). The attributes of the $M_P$ metric are derived from a direct or indirect evaluation of the users. However, we attempt to assess the quality of questions using objective quality metrics. While \citet{UAC-Ponzanelli-2014} attempted to improve the review process, our goal is to examine whether the subjective evaluation of SO questions/answers agrees with objective evaluation or not. We measure ten objective quality aspects (e.g., text readability, topic entropy) and build five machine learning models using them to predict promoted and discouraged questions. Our models significantly outperform the models by \citet{UAC-Ponzanelli-2014}. \citet{AEA-Calefato-2019} conduct an empirical assessment to predict the best answers in technical Q\&A sites. While they attempt to assess the quality of answers, we contrast the subjective quality control mechanism of SO with objective quality metrics.


Several studies \cite{ARI-Smith-1967, ANR-Flesh-1948, ACR-Coleman-1975, SGA-Mc-1969, TTO-Gunning-1952, ANR-Flesh-1948} measure the text readability as a part of objective quality assessment. In this study, we also compute the text readability of the question description to assess the reading ease of description texts. 
Besides text readability, we measure the readability of the code snippets included in the SO questions as suggested by many studies \cite{LAM-Buse-2010, MRT-Daka-2015, ASM-Posnett-2011, ICR-Scalabrino-2016,AMF-Buse-2008, USO-Treude-2017}. Code snippets included in SO are often incomplete and noisy, so the readability is found low for both question classes. We also measure the understandability of the code snippets as existing studies suggest \cite{AAC-Trockman-2018,EOS-Lin-2008, AAC-Scalabrino-2017}. \citet{AES-Zhang-2019} report that a large number of APIs hurt the code understandability. We thus count the number of APIs used in the code snippet to estimate the understandability of code snippets. 

Several recent studies \cite{FQT-Yang-2016, GET-Horton-2018} focus on assessing usability (e.g., parsability) of the code snippets of Q\&A sites. \citet{FQT-Yang-2016} analyze the usability of 914,974 Java code snippets from SO and report that only 3.89\% are parsable. They analyze the code snippets found in only the accepted answers of SO. However, we investigated the parsability of the code snippets included in the question body to estimate their reusability. Although we found a higher parsing rate than that of \citet{FQT-Yang-2016}, the parsing rates of code snippets found in promoted or discouraged questions are comparable.  


Several studies \cite{UTF-Wang-2018, ABO-Squire-2014, MSA-Calefato-2015, HTA-Calefato-2018} attempt to understand the factors (e.g., information presentation, question posting time) behind fast answers, high-quality posts, or successful answers. We also find several metrics that affect the questions' quality. For example, users often provide downvotes when the topics of the questions are less specific or ambiguous. 
In addition, several recent works \cite{SAF-Lin-2018, SPD-Calefato-2018, TDT-Novielli-2014, ABS-Novielli-2018, LAS-Islam-2017} focus on identifying developers’ emotions (e.g., positive, negative) embedded within the communication traces left in collaborative Q\&A sites like SO. \citet{TDT-Novielli-2014} claimed that the emotional style of a technical question does influence the chance of promptly getting a satisfying answer. However, we investigate whether sentiment polarity influences the quality of the SO questions. Here, the sentiment polarity does not have a noticeable impact on the question quality.

\section{Threats to validity}
\label{threatToValidity}

Threats to \emph{external validity} relate to the generalizability of results. We analyze about 2.5 million SO questions. However, the approaches we tried might show results variation when applied to different types of programming problems.  To mitigate this issue, we chose to include questions related to four different programming languages - Java, C\#, JavaScript, and Python in our dataset. Two of them are dynamically typed languages, and the remaining two are statically typed languages. 
An evaluation of our approach that involves questions from four different languages could measure the effect of this threat. The question quality classification models might be sensitive to our dataset. Therefore, we build five different classification models and analyze their performance individually.


Threats to \emph{internal validity} relate to experimental errors and biases \cite{ACO-Tian-2014}. The quality assessment of the SO questions is threatened by the evaluation systems. Using existing tools and approaches, we assess the objective quality of all of the components (e.g., title, body, tags) of questions. We then investigate how the objective evaluation aligns with the subjective evaluation from SO. However, we manually investigate the questions when the results seem inconsistent. This mixed approach mitigates this threat. 
Nevertheless, we agree that our metrics might not be the best, and improving each could change the results. Therefore, our study can be further extended by capturing more appropriate metrics or conducting a follow-up manual analysis.

\emph{Statistical Conclusion} threats concern the fact that the data is sufficient to support the claims. We considered statistically significant samples in all our result analyses. We relied on the crowd assessment to collect data, not on the manual inspection of questions that made possible to gather enough data.

Another issue is that the promoted and discouraged question samples are imbalanced. The number of promoted samples is higher than the discouraged sample for all the programming problems. To resolve this imbalanced situation, we undersampled promoted questions to balance the dataset. Then we analyze the results for both the balanced and imbalanced datasets. Thus we mitigate this issue.

\section{Conclusion}
\label{conclusion}

Q\&A websites like SO have offered a new means for programmers to participate in social learning. The quality of SO posts is assessed by users through voting. To keep the evaluation process flawless and reliable, we attempt to investigate the agreement between the subjective evaluation of questions by the users and objective quality assessment. We use ten objective metrics to capture all the key aspects of the objective quality of a SO question.
We find that topic entropy, metric entropy, text-code ratio, and text-code correlation completely agree with the subjective evaluation. Reusability (i.e., parsability) of the code snippets and text readability completely disagree with the subjective evaluation. However, code understandability either agree or disagree based on the programming languages, whereas title quality, code readability, and sentiment polarity have almost equal outcome for both the promoted and discouraged questions. Our investigation might provide an initial benchmark for further agreement analysis between subjectivity and objectivity.

Understanding and classifying question quality is essential to maintain a good user experience of Q\&A services. Low-quality questions are regarded as less useful for online Q\&A sites. We developed five different machine learning models to classify the promoted and discouraged questions. Our machine learning models can classify the questions with a maximum of about 76\%-87\% accuracy, which is significantly higher than state-of-the-art models.

\smallskip
\noindent \textbf{Acknowledgment:} This research is supported in part by the Natural Sciences and Engineering Research Council of Canada (NSERC) Discovery grants, and by an NSERC Collaborative Research and Training Experience  (CREATE) grant, and by two Canada First Research Excellence Fund (CFREF) grants coordinated by the Global Institute for Food Security (GIFS) and the Global Institute for Water Security (GIWS).


\balance



\bibliographystyle{plainnat}

\small{
    \bibliography{bibtex.bib}
}

\end{document}